\documentclass{article}
\usepackage[utf8]{inputenc}
\usepackage{caption}
\usepackage{subcaption}
\usepackage{multicol}
\usepackage{graphicx}
\usepackage{amsmath}
\usepackage[a4paper]{geometry}
 \usepackage{cite}
   \usepackage{setspace}

\newcommand{\ppt}{$p_{\rm T}$}

\usepackage{graphicx}

\usepackage{lineno}
\providecommand{\keywords}[1]{\textbf{\textit{Keywords---}} #1}
\usepackage{lineno}

\usepackage{graphicx}
\usepackage{multicol}
 \usepackage{float}
\usepackage{lineno}
\begin{document}
\begin{center}
{\Large \bf Pseudorapidity dependence of the {\ppt} spectra of
charged hadrons in $pp$ collisions at $\sqrt{s}$~= 0.9 and 2.36 TeV}

\vskip1.0cm Pei-Pin~Yang$^{1,}${\footnote{emails:
peipinyangshanxi@163.com; yangpeipin@qq.com (P.-P. Yang)}},
M.~Ajaz$^{2,}${\footnote{Corresponding author  emails:
ajaz@awkum.edu.pk; muhammad.ajaz@cern.ch (M. Ajaz)}},
M.~Waqas$^{3,}${\footnote{Corresponding author. emails:
waqas\_phy313@yahoo.com; waqas\_phy313@ucas.ac.cn (M. Waqas)}},
Fu-Hu~Liu$^{1,}${\footnote{Corresponding author. emails:
fuhuliu@163.com; fuhuliu@sxu.edu.cn (F.-H. Liu)}},
M.~K.~Suleymanov$^{4,}${\footnote{email:
mais.suleymanov@bsu.edu.az (M. K. Suleymanov)}}

{\small\it $^1$ Institute of Theoretical Physics, State Key
Laboratory of Quantum Optics and Quantum Optics Devices \& \\
Collaborative Innovation Center of Extreme Optics, Shanxi
University, Taiyuan 030006,
China \\
$^2$ Department of Physics, Abdul Wali Khan University Mardan, 23200 Mardan, Pakistan\\
$^3$ School of Nuclear Science and Technology, University of Chinese Academy of Sciences, Beijing 100049, China\\
$^4$ Department of Physics, Baku State University, Baku,
Azerbaijan
 }\\
\end{center}

\vskip1.0cm
%%%%%%%%%%%%%%%%%%%%%%
%\linenumbers
%\date{October 2021}

%\begin{document}

%\maketitle
\begin{abstract}
We report the predictions of different Monte Carlo event
generators including HIJING, Pythia, and QGSJETII in comparison
with the experimental data measured by the CMS Collaboration at
CERN in proton-proton ($pp$) collisions at center-of-mass energy
$\sqrt{s}$~= 0.9 and 2.36 TeV. The CMS experimental transverse
momentum ({\ppt} or $p_{\perp}$) spectra of charged hadrons were measured for pseudorapidity range 0 $\le$ $\eta$ $\le$ 2.4 with bin width of $\eta$ = 0.2 (for {\ppt} from 0.1 to 2 GeV/$c$) and a single bin of $\eta$ for $\lvert$$\eta$$\rvert$ $<$ 2.4 (for {\ppt} from 0.1 to 4 GeV/$c$). Pythia reproduced the {\ppt} spectra with reasonable agreement for most of the {\ppt} range. It depicts better results in the case of the $|\eta|<$ 2.4 than HIJING and QGSJETII which could reproduce the spectra in a limited {\ppt} range. Furthermore, to analyze the {\ppt} spectra of charged hadrons measured by the CMS Collaboration, we used a three-component function (structured from the Boltzmann distribution) and the $q$-dual function (from the
$q$-dual statistics) to extract parameter values relevant for the
study of bulk properties of hadronic matter at high energy. We have
also applied the two analytic functions over the model
predictions. The values extracted by the functions from the HIJING and Pythia models are closer to the experimental data than the QGSJETII model. Although the models could reproduce the $p_T$ spectra of all charged particles in some of the $p_T$ range but none of them could reproduce the distributions over the entire $p_T$ range and in all the pseudorapidity regions.

\vskip0.5cm \keywords{Transverse momentum spectra, pseudorapidity
distribution, Monte Carlo prediction, LHC energies, proton-proton
collisions}
\\


\end{abstract}

%\maketitle

\vskip1.0cm
\begin{multicols}{2}
\section{Introduction}\label{sec1}

In the laboratory, the strongly interacting matter can only be produced
by colliding nuclei at high energies. The matter passes through different phases and finally, free hadrons are detected by the detectors. The study of particle collisions itself is a complex phenomenon that makes it difficult to understand the geometry and measure global or bulk properties of the interactions. Besides experimental measurements, it is therefore
important to use theoretical models or analytic functions to
deduce different parameters such as the effective temperature
($T$), kinetic freeze-out temperature ($T_0$), average transverse
flow velocity ($\langle\beta_T\rangle$ or $\beta_T$ in simple
term) and kinetic freeze-out volume ($V$) \cite{1, 2, 3, 4} based
on experimental data. Here, $T$ contains the contribution of
thermal motion reflected by $T_0$ and transverse flow effect
reflected by $\beta_T$. $T_0$, $\beta_T$ and $V$ are quantities at
the stage of kinetic freeze-out.

The RHIC at BNL~\cite{5, 6} and LHC at CERN ~\cite{7} used proton-proton ($pp$) and proton-nucleus ($pA$) collisions as well
as heavy-ion or nucleus-nucleus ($AA$) collisions at high energies
to study the bulk properties of the strongly interacting
matter~\cite{8}. Simultaneously, various Monte Carlo models (event generators) and analytic functions have
been proposed to describe the stages of evolution of the system~\cite{9, 10}. These analyses are hopeful to
extract some parameters and then to understand the mechanisms and
laws of multi-particle production.

Measurements of pseudorapidity ($\eta$) distributions can provide
important constraints in the simulation of $pp$ collision
characteristics such as the angle at which the particle is emitted to the beam axis \cite{11}. It is given by $\eta \equiv -\ln\tan(\theta/2)$, where $\theta$ is the emission angle of the particle with respect to the beam axis. In addition, $\eta$
is an approximate expression of the rapidity, $y \equiv (1/2)\ln[(E+p_z)/(E-p_z)]$, at high energy. Here $E$ and $p_z$ denote the energy and longitudinal momentum respectively. Both
$\eta$ and $y$ can reflect the degree of longitudinal extension of
the collision system. Measurements of transverse momentum ({\ppt}
or $p_{\perp}$) can provide important information on the transverse excitation degree of the collisions system. In particular, the {\ppt} spectra in a given $\eta$ range can provide
the information of $\eta$ dependent transverse excitation degree. Here the transverse excitation degree is described by various types of temperatures.

One may use some MC models or analytic functions to analyze the
experimental data and extract some parameters. Generally, the MC
model results are compared with the experimental data to find the
mechanisms of particle production. Meanwhile, the analytic
functions are used to fit the experimental data to find the laws
obeyed by the multiple particles. It is interesting for us to fit
the experimental data and the MC model results by the analytic
functions simultaneously, and to extract the related parameters
and compare them. This is the goal of the paper. Three MC models,
HIJING~\cite{hijing}, Pythia~\cite{pythia} and
QGSJETII~\cite{qgsjetII1,qgsjetII2,qgsjetII3,qgsjetII4}, and two analytic functions, a
three-component function structured from the Boltzmann
distribution (statistics) and the $q$-dual function from the $q$-dual statistics, are used in the paper to compare with the experimental data measured by the CMS Collaboration~\cite{cms}. Meanwhile, the two analytic functions are also compared with the three MC models.

The CMS experimental data are measured in given ranges of kinetic
quantities. Due to the ranges, we study the transverse momentum
{\ppt} spectra, $dN_{ch}/dp_{\rm T}$, of charged hadrons from
{\ppt} = 0.1 to 2 GeV/$c$ in small bins of $\eta$ with a gap of
0.2 units from $|\eta|=0$ to 2.4. Meanwhile, the {\ppt} spectra is
studied from {\ppt} = 0.1 to 4 GeV/$c$, in a $\eta$ range of
$|\eta|<$ 2.4. In addition, it is worth mentioning that in the
case of experimental data, the daughter particles of those parents
that decay with less than 1 cm$/c$ life-time are included in the
charged hadrons, whereas other secondary particle products are not
included in the definition of charged hadrons~\cite{cms}. This
experimental treatment minimizes the misjudgment rate.

The rest of the paper is presented as follows: Section~\ref{sec2}
is dedicated to the method and brief description of the MC models
used for comparison with the experimental measurements as well as
the analytic functions used to fit the data. It is followed by
Section~\ref{sec3}, where a complete description of the analysis
procedure along with the simulation results using
HIJING~\cite{hijing}, Pythia~\cite{pythia} and
QGSJETII~\cite{qgsjetII1} in comparison with CMS experimental
data~\cite{cms} is presented. Furthermore, fitting the experimental data as well as model predictions with analytic functions are also presented. The work is summarized in Section~\ref{sec4} where conclusions are drawn.

\section{Method and models}\label{sec2}

\subsection{Three MC models}

First of all, we introduce the general cases in the comparative
study of charged hadron spectra calculated from three MC models
and measured by the CMS collaboration~\cite{cms} in $pp$
collisions at $\sqrt{s}$~= 0.9 and 2.36 TeV. The spectra in the
experimental measurements are normalized to all
non-single-diffractive (NSD) events where corrections for
branching ratios, acceptance, trigger, and selection efficiency are used. The same initial condition as the measurements are
employed in the simulation data. Three MC model predictions are used for comparison with the experimental data. Here, {\ppt}
spectra from 0.1 to 2 GeV/$c$ in twelve equal bins of $\eta$ from
0 to 2.4 and {\ppt} spectra from 0.1 to 4 GeV$c$ for $|\eta|<2.4$
are used in simulation following the data. One million events are simulated in each case for comparison to the experimental data. This renders the statistical errors for MC models are very small and mostly not visible. A brief description of the MC models used in the study is given below.

{\bf HIJING} model -- a Monte Carlo event generator and an acronym
for heavy-ion jet interaction generator. For jet fragmentation,
the model uses the pQCD (perturbative quantum chromodynamics)
inspired models including Dual Parton model (DPM)~\cite{23} and
Lund string fragmentation model~\cite{string} of multi-string
phenomenology at low and medium energies to produce effects of
soft interactions. The model is particularly developed to study
the mini-jets and associated particle production, soft excitation,
distribution functions of parton, and study of interactions of jets
in the dense medium~\cite{25,26}. The model is using both
Pythia~5.3 and Jetset~7.2 for the generation of kinematic
variables and jet fragmentation respectively~\cite{27}. The aim of
the model development is to explore the range of
initial conditions appearing in the case of ultra-relativistic
heavy-ion interactions. The HIJING model is designed to simulate multi-particle production in different interaction systems of $pp$,
$pA$ and $AA$ at high energies. The model has the capacity to reproduce the parameters of the data from $pp$ collisions for the energy range of $\sqrt{s}=5$--2000~GeV~\cite{25}. The HIJING model reports a wider range of phenomenological problems regarding high-energy nuclear interactions such as the production of multi-particles in different types of high-energy interactions, a study of mini-jets and jets production. HIJING was the only MC generator, when developed, to comprise the pQCD methodology of Pythia to multiple-jet processes and other processes like nuclear jet quenching and parton shadowing. In HIJING different types of
high energy interactions are based on Glauber geometry. The model is now rewritten in C++ \cite{hijing++} with many advanced features using all models from from its predecessor.

{\bf Pythia} model -- a Monte Carlo event generator based on the
Lund string fragmentation~\cite{string} for particle production.
It is the most widely used MC simulation package with emphasis for
$pp$ collisions at high energies~\cite{pythia}. The recent
versions of the Pythia model viz Pythia8 can produce high energy heavy-ion exclusive hadronic collisions in the final
states. The Pythia8 model provides a good description of the
minimum bias as well as the hard-scale $pp$ collisions which are
further extended to describe the interactions in heavy-ion
collisions. It is an event generator that can successfully describe the collisions of high-energy protons, electrons, photons,
and heavy nuclei. The model can explain soft and hard
interactions, final- and initial-state parton showers,
distribution of partons, decays, fragmentation and multi-parton
interactions.

{\bf QGSJET} model -- a Monte Carlo event generator based on the
Quark-Gluon String model (QGS) developed for hadron-hadron
interactions~\cite{qgsjetII1}. Besides the QGS model, it is also
using the Gribov-Regge theory and concept from the theoretical
framework of the LUND algorithm to observe multiple-scattering phenomena. The recent version QGSJETII-04 is also tuned with the
LHC data. The advantage of this model over the others is that it
has the least number of free parameters, and hence, therefore, tuning of only a few parameters is required for full
implementation of multi-pomeron exchanges and interactions.
Before going into description of the second part of the analyses by two analytic function, it is good to mention that similar work of models' prediction in comparison of experimental data can be found in our previous work ~\cite{ajaz1,ajaz2,Khan,ajaz3,Li,universe}. 
\subsection{Two analytic functions}

Then secondly, for the comparison of analytic functions with the
experimental data and model results, we use two different
functions. The first one is a three-component function which is
structured from the Boltzmann distribution (statistics). The
second one is the $q$-dual function from the $q$-dual statistics.

One has the Boltzmann distribution with the effective temperature
$T_B$ to be
\begin{align}
\frac{1}{2\pi p_{\rm T}}\frac{d^2N}{dydp_{\rm{T}}} =&\,\frac{gV}{(2\pi)^3} m_{\rm{T}}\cosh y \times \nonumber\\
&\exp\left(-\frac{m_{\rm{T}}\cosh y-\mu}{T_B}\right), \label{eq1}
\end{align}
where $N$ is the number of particles, $g$ is the degeneracy
factor, $V$ is the volume, $m_{\rm T}=\sqrt{p_{\rm T}^2+m_0^2}$ is
the transverse mass, $m_0$ is the rest mass and $\mu$ is the
chemical potential. To structure the three-component function
conveniently, one needs the probability density function at
mid-rapidity. For the $i$-th component, we have
\begin{align}
f_i(p_{\rm{T}}) = \frac{1}{N}\frac{dN}{dp_{\rm T}}=
A_ip_{\rm{T}}m_{\rm{T}}\exp\left(-\frac{m_{\rm{T}}-\mu}{T_i}\right),
\label{eq2}
\end{align}
where $A_i$ is the normalization constant which makes
$\int_0^{\infty} f_i(p_{\rm{T}})dp_{\rm T}=1$. At high energy, one
has $\mu\approx0$.

The three-component function is
\begin{align}
f(p_{\rm{T}})
=k_{1}f_{1}(p_{\rm{T}})+k_{2}f_{2}(p_{\rm{T}})+(1-k_{1}-k_{2})f_{3}(p_{\rm{T}}),
\label{eq3}
\end{align}
where $k_1$, $k_2$ and $1-k_1-k_2$ are the fractions contributed
by the first,  second and third components. The average value of
$T_i$ obtained by weighting $k_i$ for $T_i$ is
\begin{align}
T_B =k_1T_1+k_2T_2+(1-k_1-k_2)T_3, \label{eq4}
\end{align}
where the denominator is 1 which is not displayed.

We have used the three-component function because the Boltzmann
distribution and its two-component function do not work well. In
the three-component function, the parameters include the free
parameters $T_1$, $T_2$, $T_3$, $k_1$ and $k_2$ and the
normalization constant $N_0$. To avoid repetition, $T_B$ is shown
usually to represent the five free parameters in the analysis.

One may also structure the three-component function from the
invariant yield [Eq. (1)] directly, in which the three expressions
are added directly and the parameters include the free parameters
$T_1$, $T_2$ and $T_3$ and the normalization constants $V_1$,
$V_2$ and $V_3$. One has $T_B=\sum_i k_iT_i$ and $V=\sum_i V_i$,
where $k_i=V_i/V$ and $k_i$ is also the ratio of the area under
the $i$-th curve to the total area under the fit curve.

The $q$-dual function from the $q$-dual statistics is
\begin{align}
\frac{1}{2\pi p_{\rm{T}}}\frac{d^{2}N}{dydp_{\rm{T}}}=&\, \frac{gV}{(2\pi)^{3}}m_{\rm{T}}\cosh y\sum_{n=0}^{\infty}(-S)^{n}\times \nonumber\\
&\left[1-(1-q)(n+1)\frac{m_{\rm{T}}\cosh
y-\mu}{{T_{q}}}\right]^{\frac{q}{1-q}}, \label{eq5}
\end{align}
where $S=1$, $-1$ and 0 are for the Fermi-Dirac, Bose-Einstein and
Maxwell-Boltzmann statistics of particles, respectively. In the calculations, for protons and kaons, we have used S=1, and for pions, we have used S=-1. The case of S=0 is not used. In the
$q$-dual function, the parameters include the free parameters
$T_q$ and $q$ and the normalization constant $V$.

\section{Results and discussions}\label{sec3}

\begin{figure*}[]
\centering
\includegraphics[width=0.32\textwidth]{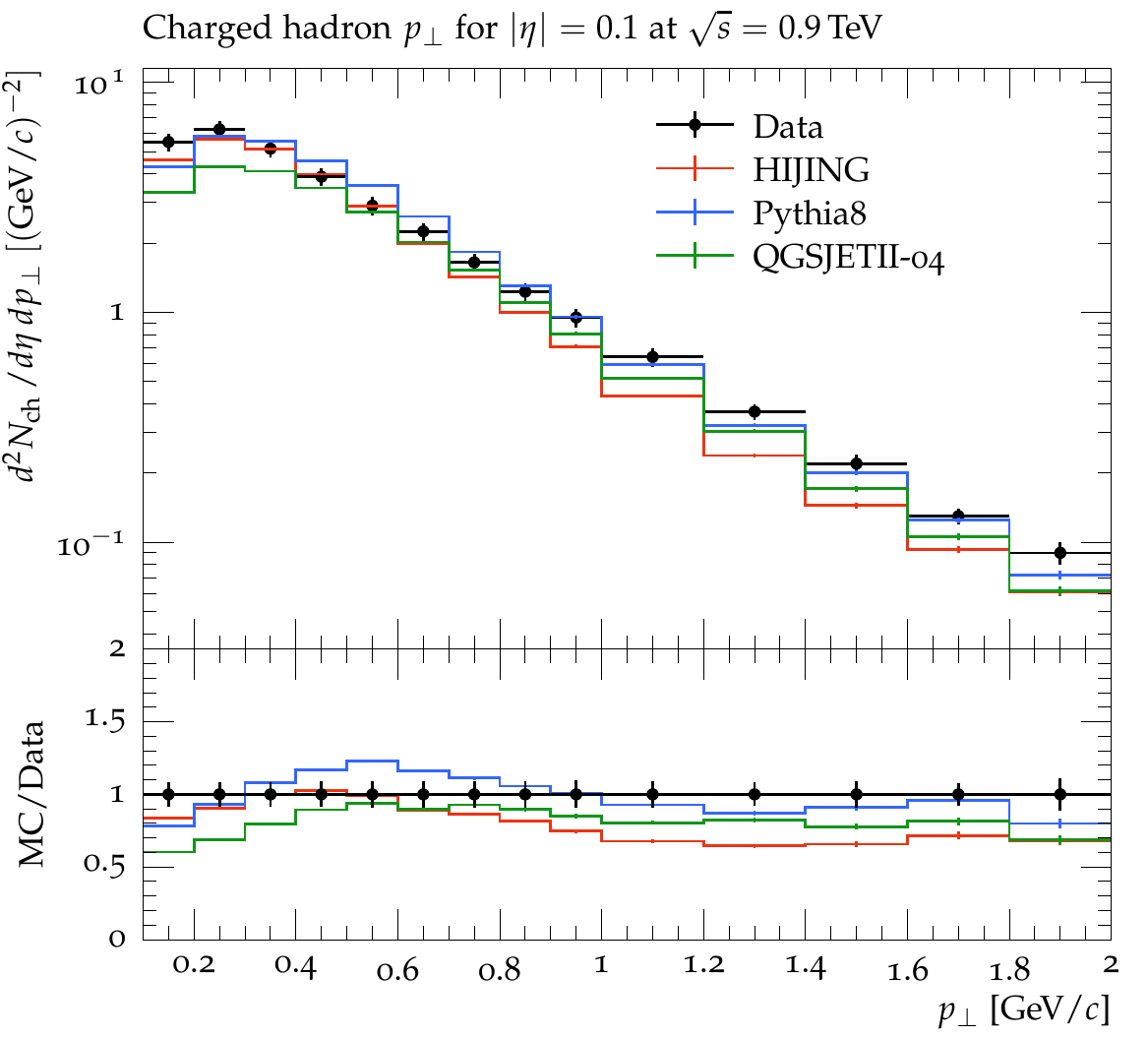}
\includegraphics[width=0.32\textwidth]{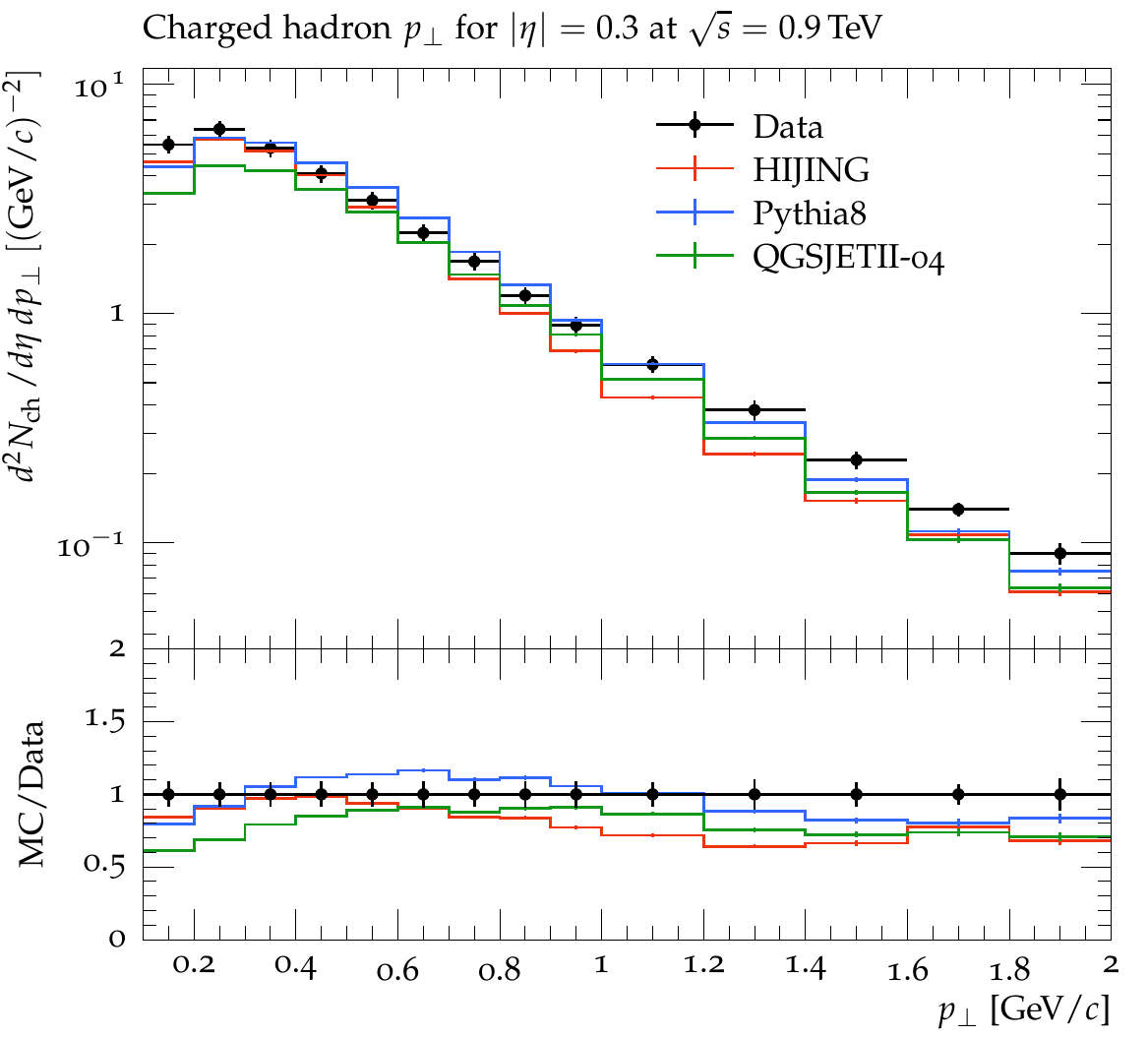}
\includegraphics[width=0.32\textwidth]{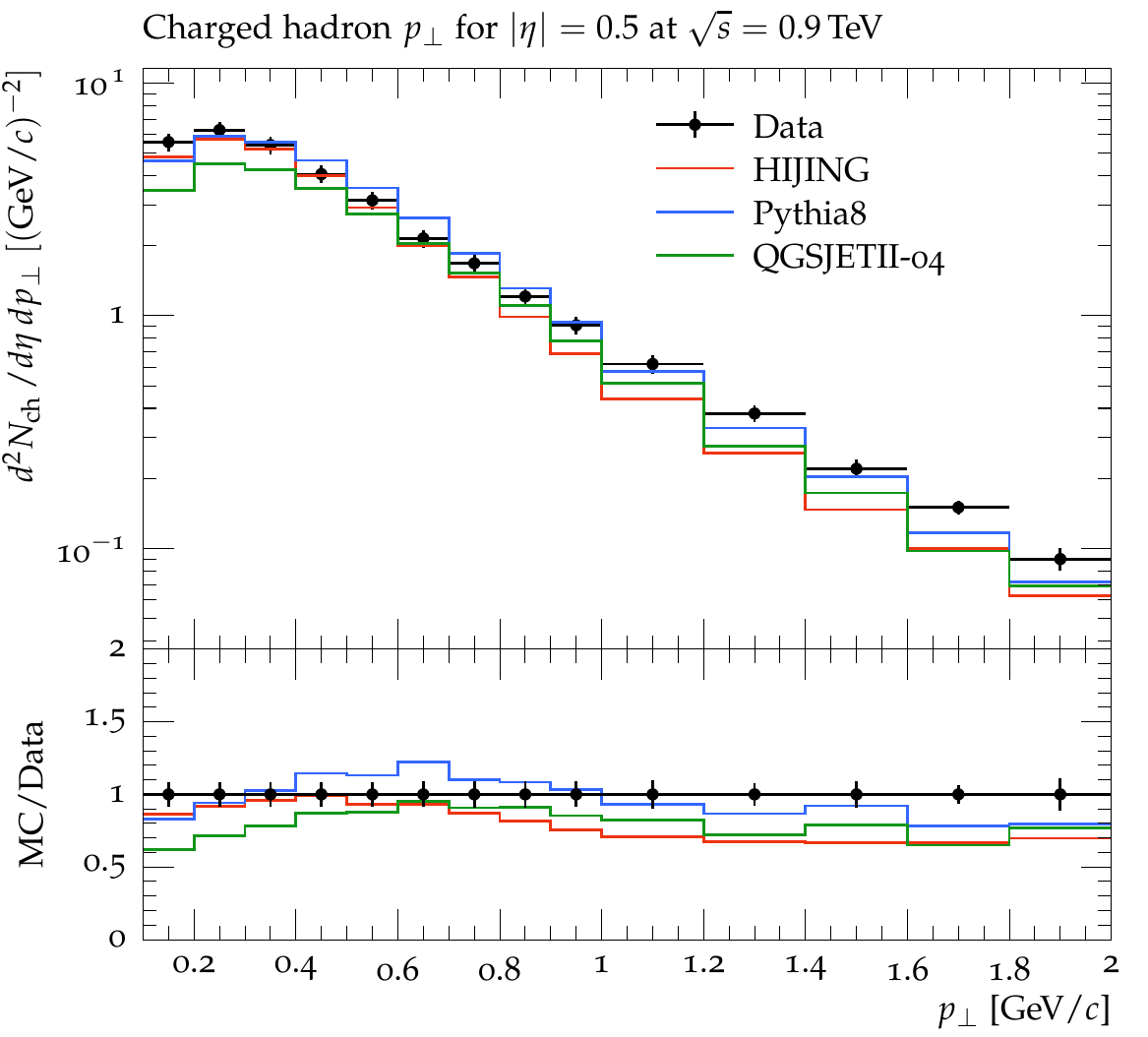}
\includegraphics[width=0.32\textwidth]{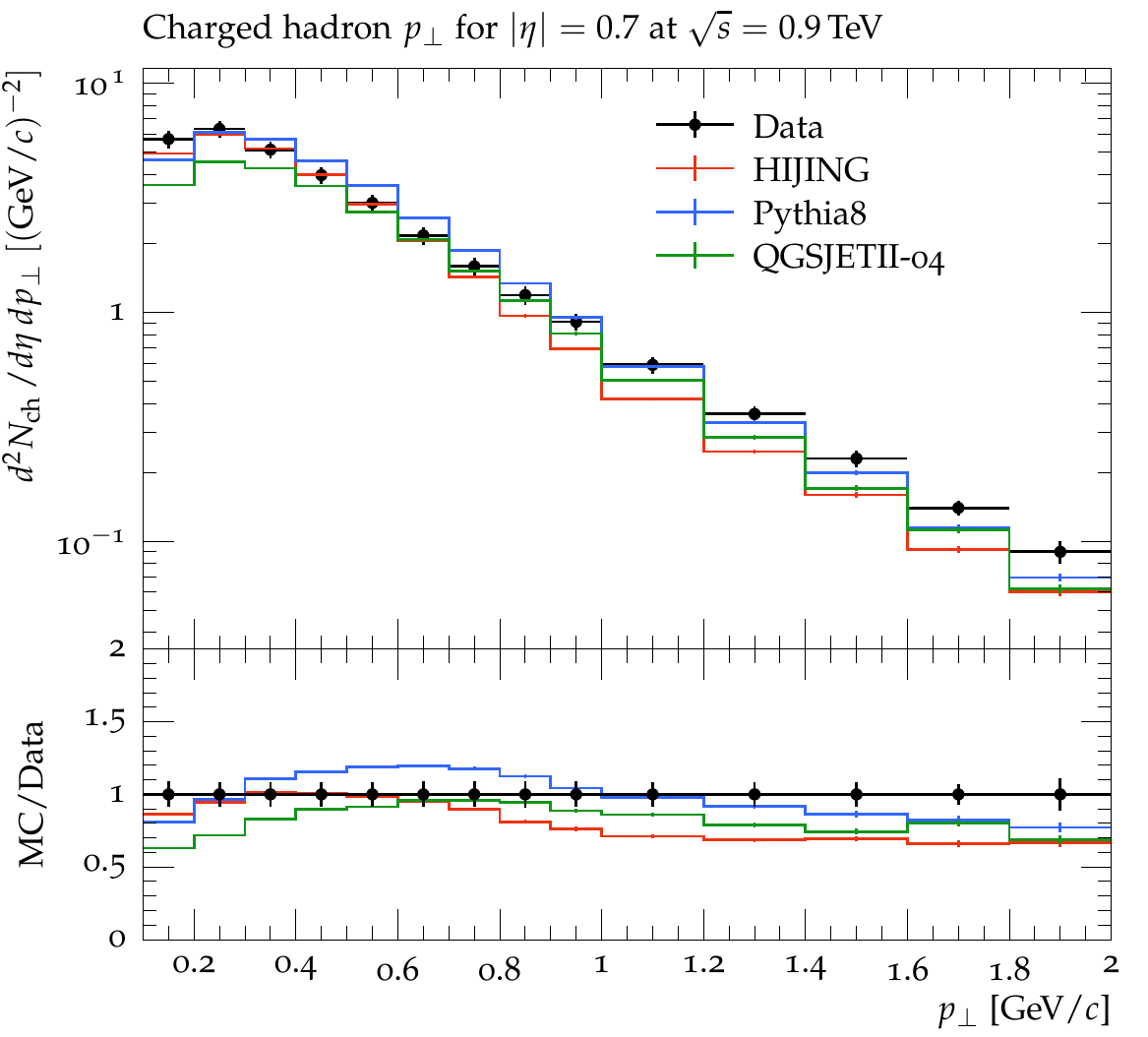}
\includegraphics[width=0.32\textwidth]{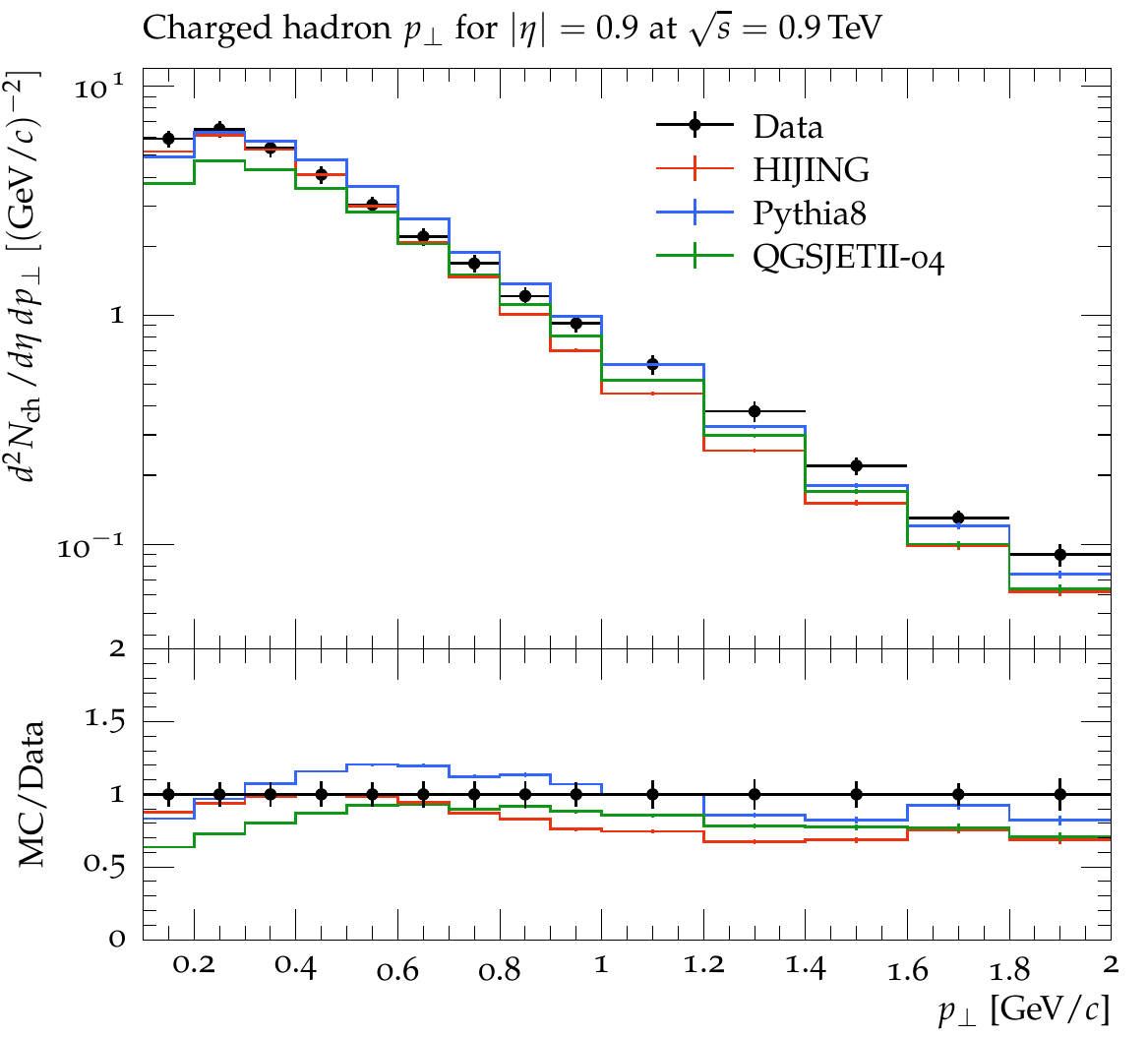}
\includegraphics[width=0.32\textwidth]{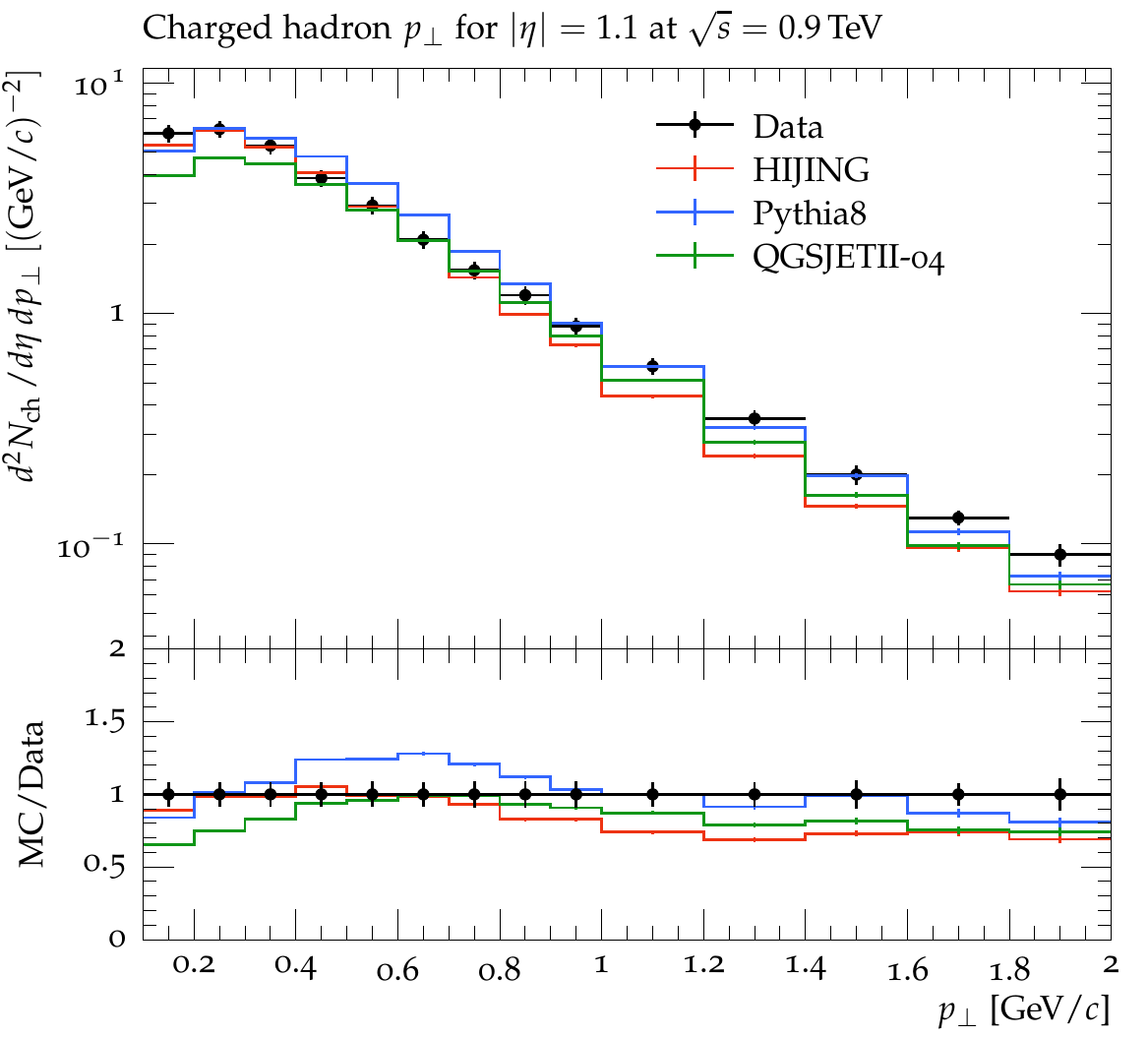}
\includegraphics[width=0.32\textwidth]{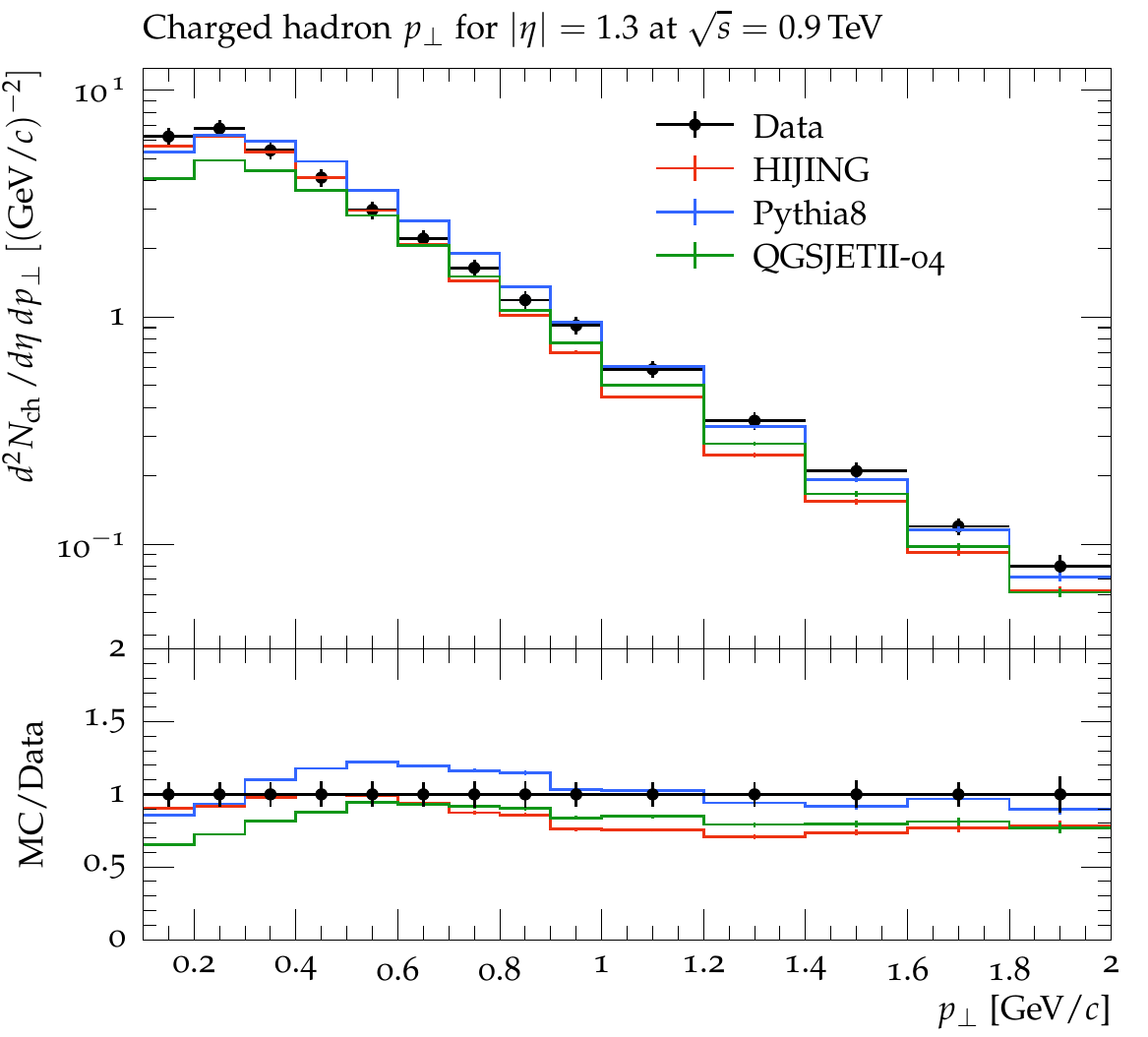}
\includegraphics[width=0.32\textwidth]{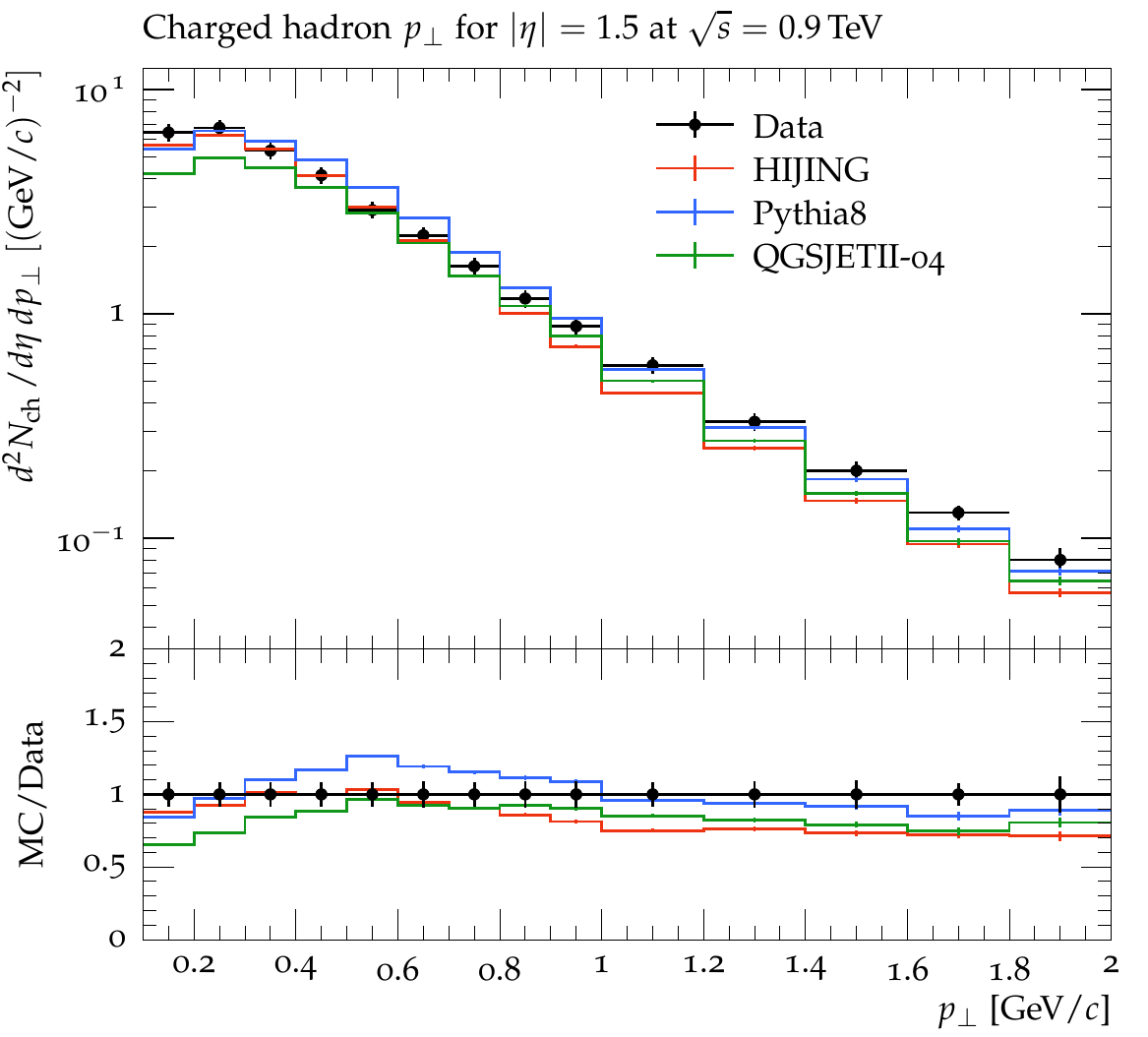}
\includegraphics[width=0.32\textwidth]{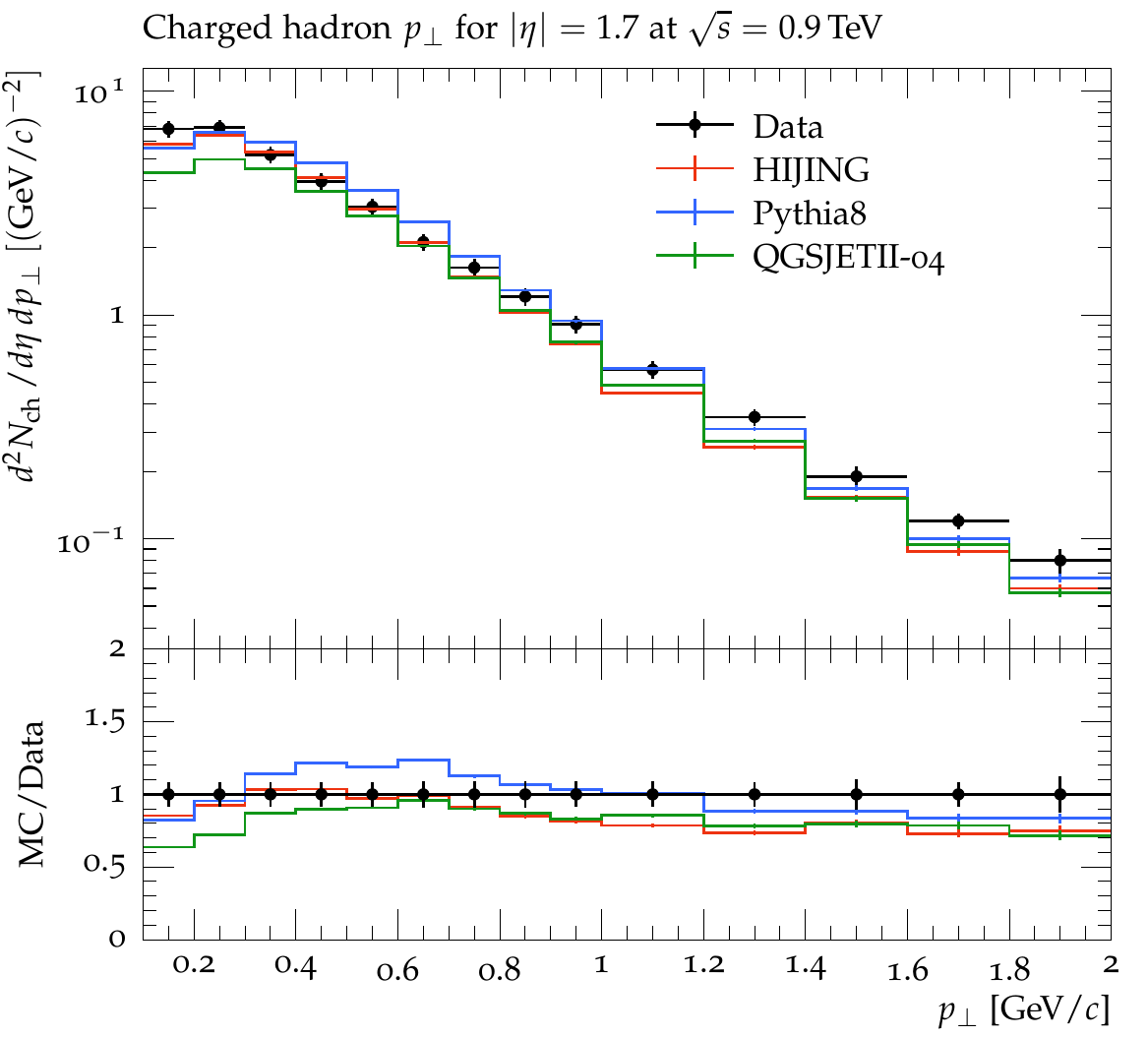}
\includegraphics[width=0.32\textwidth]{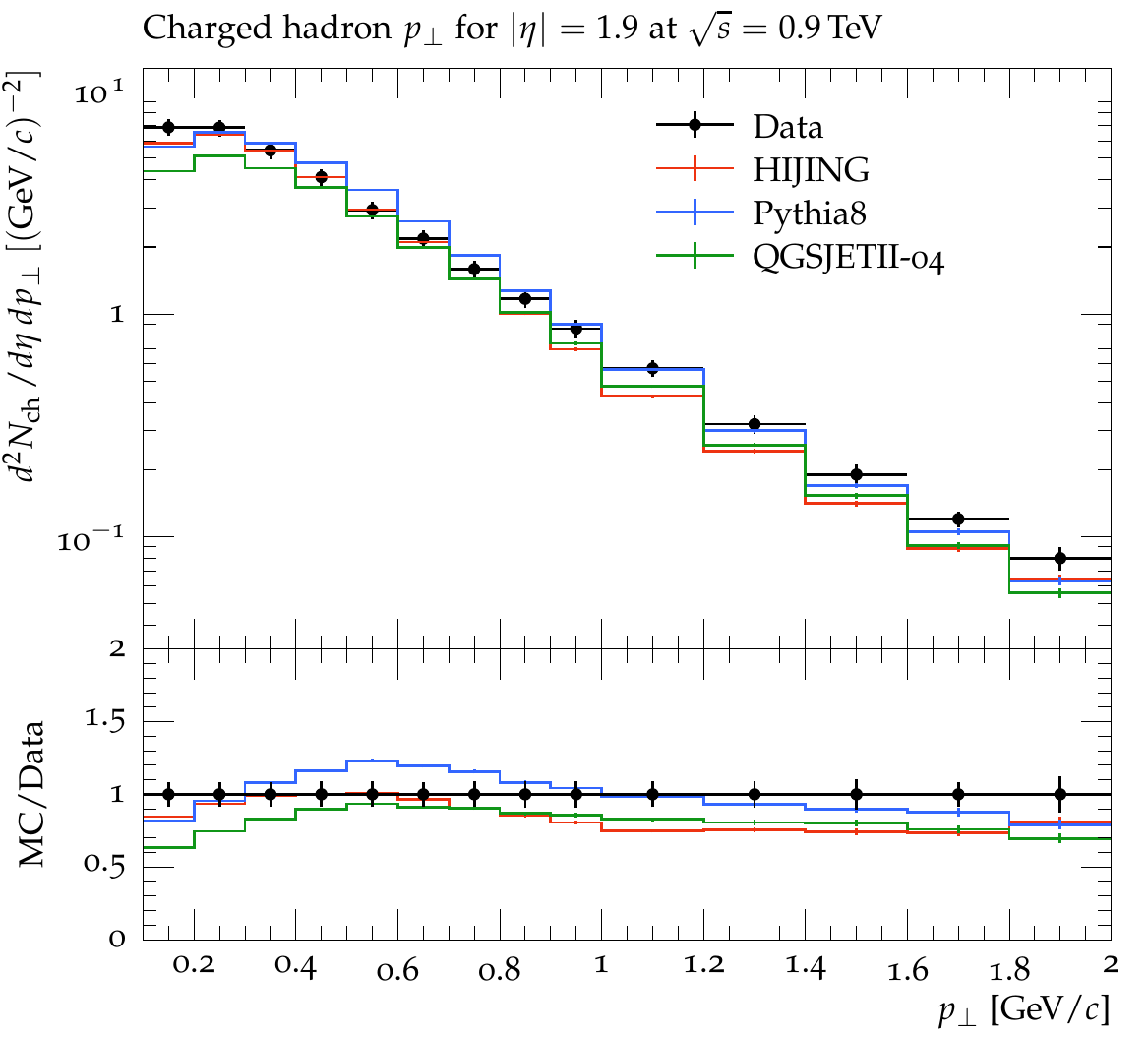}
\includegraphics[width=0.32\textwidth]{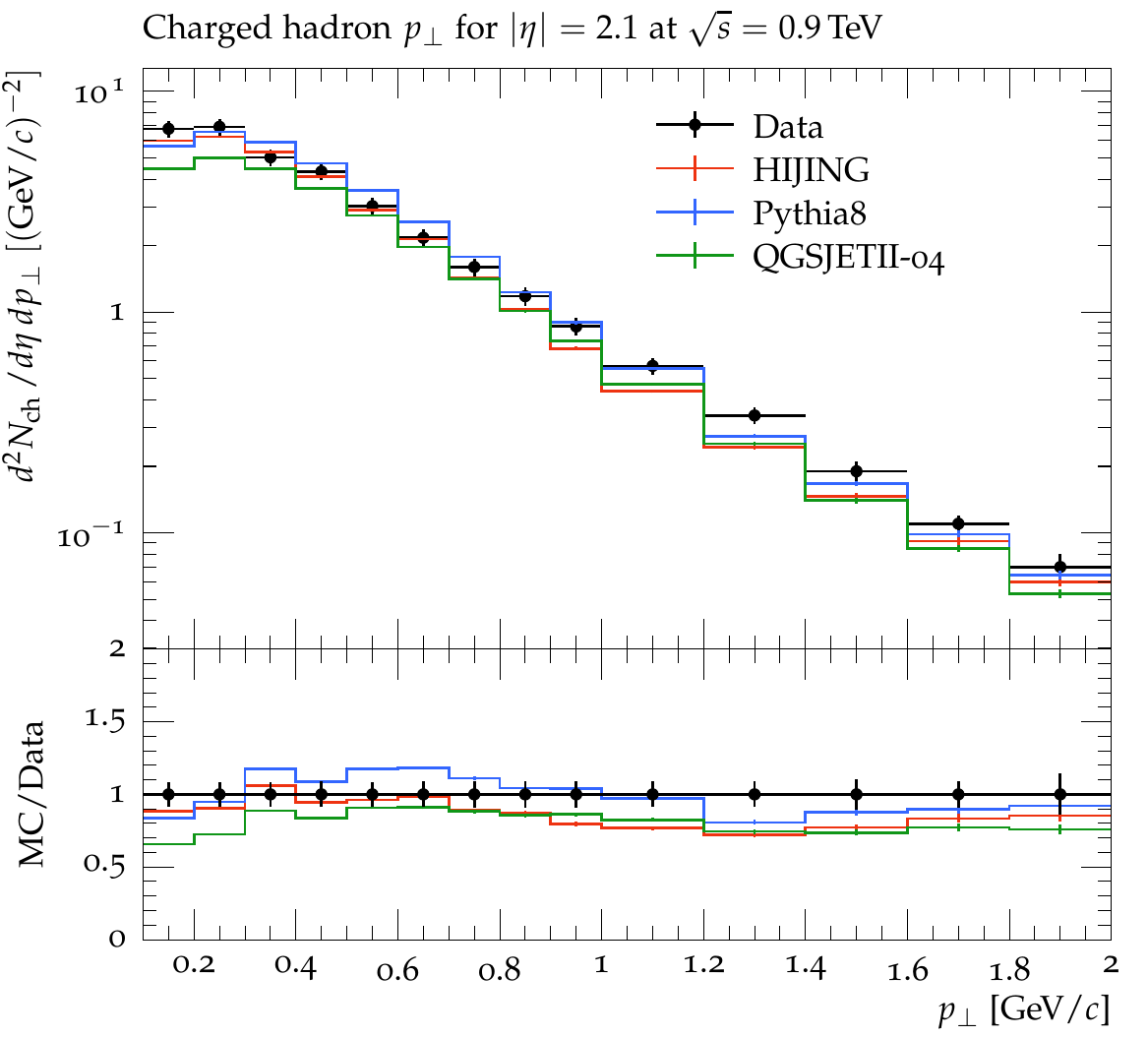}
\includegraphics[width=0.32\textwidth]{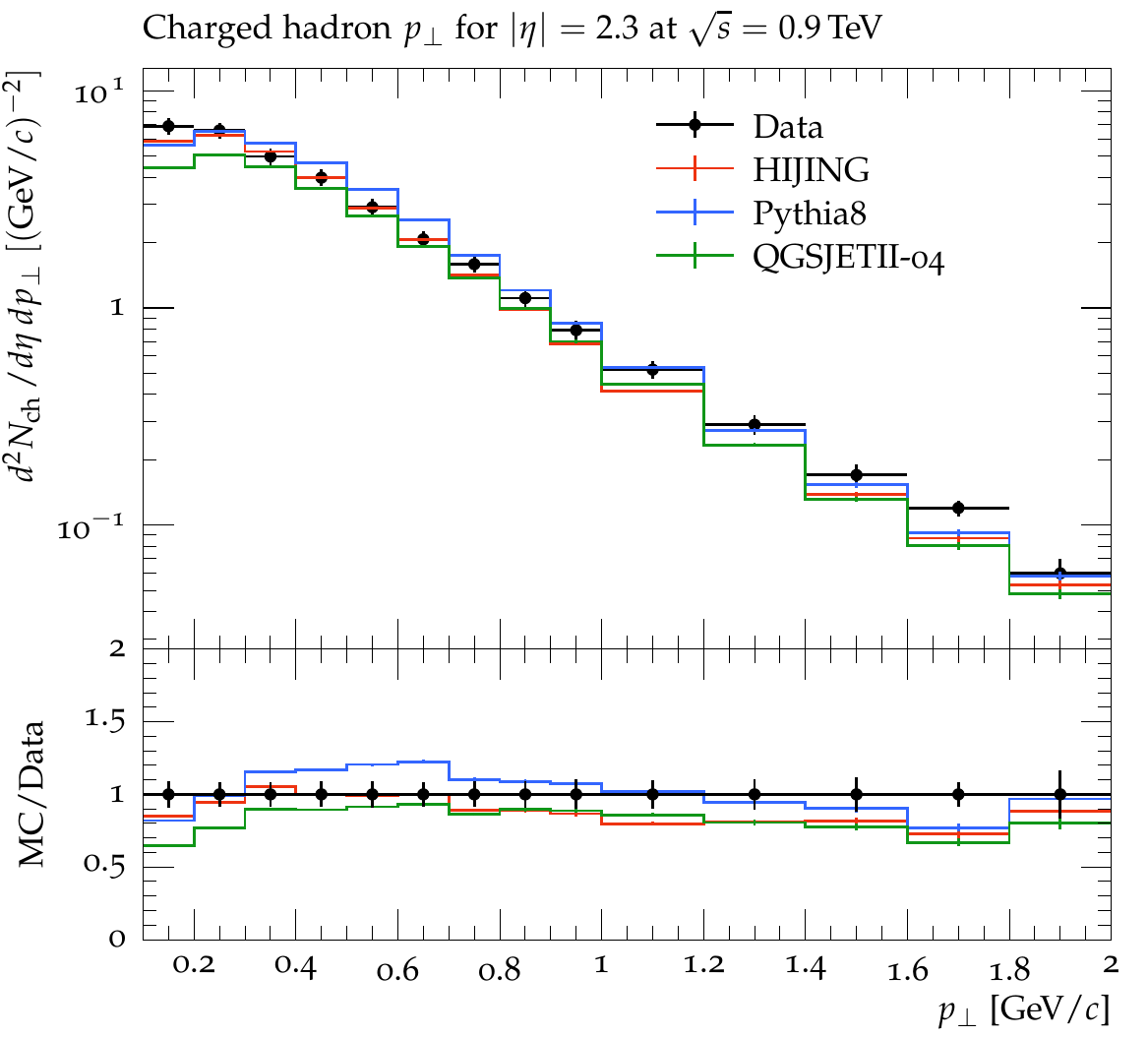}
\caption{The transverse momentum spectra shown in different $|\eta|$ bins of charged hadrons \cite{cms} are compared with the
MC model predictions \cite{hijing,string,qgsjetII1} from $pp$
collisions at $\sqrt{s}$~= 0.9 TeV. The solid black markers show the experimental data while the solid lines of different colors represent the different MC model predictions shown in the panels.}
\label{fig1}
\end{figure*}

\begin{figure*}[!]
\centering
\includegraphics[width=0.95\textwidth]{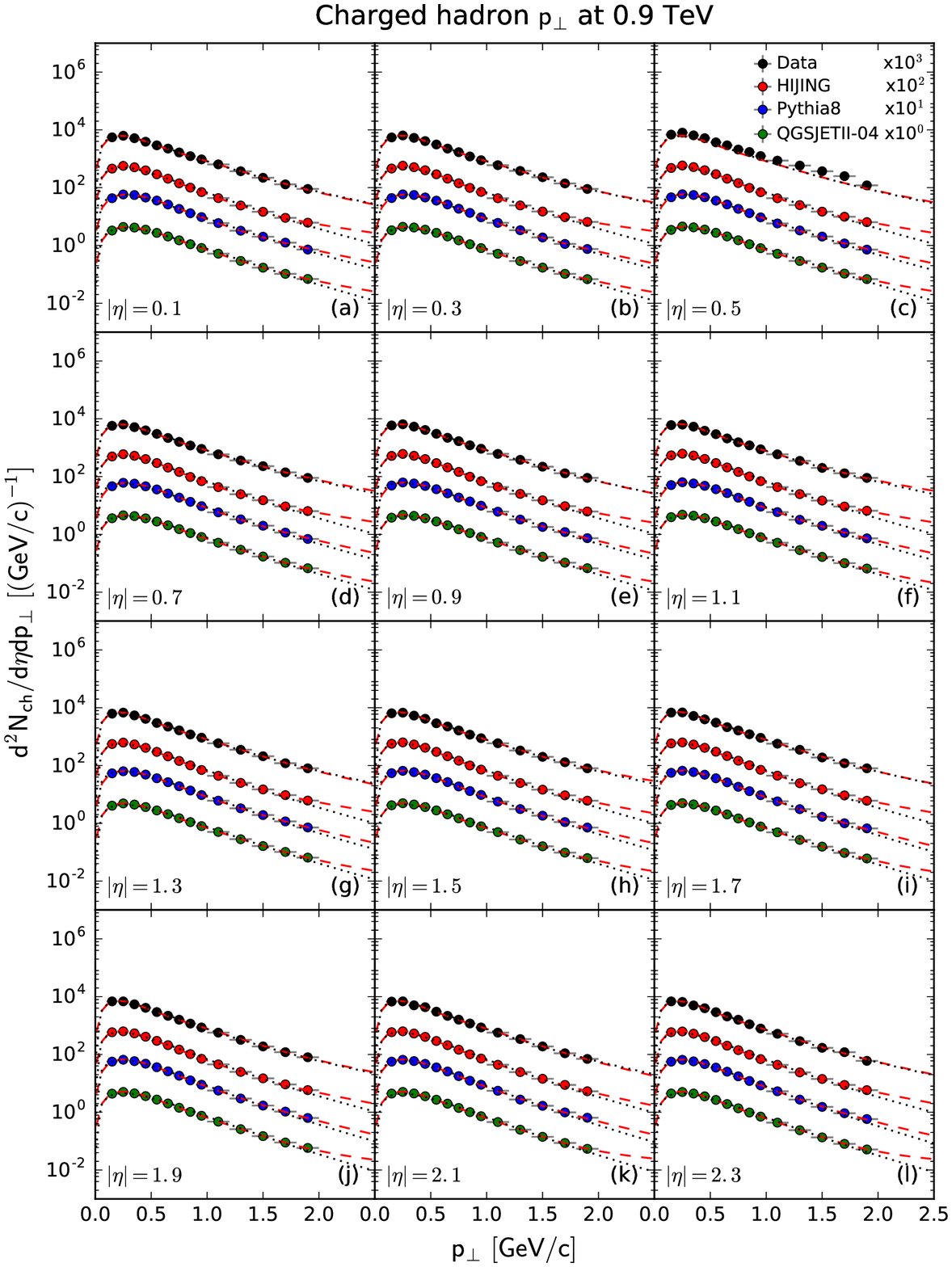}
\caption{The experimental measurements \cite{cms} as well as the
MC model predictions \cite{hijing,string,qgsjetII1} of the
transverse momentum {\ppt} spectra shown in different $|\eta|$
bins of charged hadrons in $pp$ collision at $\sqrt{s}$~= 0.9 TeV are
fitted with two analytic functions. The dashed curves represent
the fit results of the three-component function [Eq. (3)], while
the dotted curves represent the fit results of the $q$-dual
function [Eq. (5)].} \label{fig2}
\end{figure*}

The transverse momentum {\ppt} spectra of charged hadrons from
HIJING, Pythia and QGSJETII-04 models in $pp$ collisions at
$\sqrt{s}$~=0.9 TeV are presented in Figure \ref{fig1}. The
predictions of simulated results are compared with the
measurements of CMS experiment \cite{cms} in $pp$ collisions at
$\sqrt{s}=0.9$ TeV. The same cuts as the experimental data are
applied on simulations. The {\ppt} spectra of charged hadrons is
presented in small bins of $|\eta|$. From the left-up panel to the
right-down one, the values of $|\eta|$ change from 0.0--0.2 to
2.2--2.4 for {\ppt} from 0.1 to 2 GeV/$c$. The vertical error bars
in case of experimental data is sum of the quadrature addition of
systematic and statistical errors while it shows statistical
errors in case of model predictions. The horizontal line at the
data point as well as the model predictions show the size of the
{\ppt} bin. The lower panel of the graph show the ratio of the MC
predictions to the data measurements.

From Figure \ref{fig1} one can see that the HIJING model
reproduces the experimental measurements for $|\eta|$ from 0.1 to
0.7 between the {\ppt} from 0.3 to 0.7 GeV/$c$ while up to 0.8
GeV/$c$ for $0.8\le|\eta|\le2.2$ and up to 1 GeV/$c$ at
$|\eta|=2.3$. This shows that the prediction of the HIJING model
gets better with increasing the value of $|\eta|$ at $\sqrt{s}$ =
0.9 TeV. The prediction of the Pythia model has the closest
predictions amongst the other models. The model depicts the {\ppt}
spectra between 0.2--0.4, 0.8--1.2 and 1.4--1.8 GeV/$c$ for
$|\eta|=0.1$. It overestimates between 0.4--0.8 GeV/$c$ while
underestimates at 0.1, 1.3 and 2.0 GeV/$c$. The prediction of the
model has little variation in the {\ppt} spectra with increasing
$|\eta|$ and hence has similar behavior for all $|\eta|$ bins.
Finally the QGSJET model depicts the data for {\ppt} = 0.4--0.9
GeV/$c$ at $|\eta|$ = 0.1 but underestimates below 0.4 GeV/$c$ and
above 0.9 GeV/$c$. Again there is little variation in the behavior
of {\ppt} distribution of the model prediction with varying
$|\eta|$. At high {\ppt}, the prediction of QGSJET model is better
than the HIJING while at low {\ppt}, HIJING has better results.

The CMS data and the three MC results in Figure \ref{fig1} are
fitted by the three-component function [Eq. (\ref{eq3}), the
dashed curves] and the $q$-dual function [Eq. (\ref{eq5}), the
dotted curves] in Figure \ref{fig2}. Different colors of the
symbols corresponded to different cases which refer to the CMS
data or MC results marked in the panel, where different re-scaled
amounts are used in different cases. The values of $T_B$, $N_0$,
$\chi^2$ and number of degrees of freedom (ndof) are listed in
Table 1, while the values of $T_q$, $q$, $V$, $\chi^2$ and ndof
are listed in Table 2. One can see that the CMS data and the three
MC results are fitted well by the two analytic functions.

It should be noted that both the extracted temperature parameters
$T_B$ and $T_q$ are the effective temperatures which include the
contributions from both the thermal motion which is reflected by
the kinetic freeze-out temperature $T_0$ and the transverse
directed motion which is reflected by the average transverse flow
velocity $\beta_T$. There are indirect methods to decouple the
values of the aforementioned two parameters from the effective
temperature \cite{28, 29, 29a}, but it is beyond the focus of the
present work.

The values of $T_B$ in Table 1 for the data, HIJING, Pythia8 and
QGSJETll-04 are observed to decrease continuously with the
increase of $|\eta|$. Similarly in Table 2, the values of $T_q$
are initially remains unchanged in a few $|\eta|$ slices, then
decrease (remain unchanged) for the data and the MC results. But as a whole, the trend is of $T_B$ ($T_q$) gives a slight decrease with increasing $|\eta|$. This trend is due to the reason that large $|\eta|$ corresponds to large penetration between participant particles (where the large penetration means large rapidity shift) which results in smaller $T_B$ ($T_q$) and this result is consistent to our recent work \cite{30}. In Table 1, the parameter $N_0$ is the normalization constant and it shows
multiplicity, and it is observed to be approximately unchanged or
increase slightly with changing $|\eta|$. In Table 2, the entropy
based parameter $q$ is presented and it is seen that there is no
specific dependence of $q$ on $|\eta|$. Furthermore, we also
extracted the kinetic freeze-out volume $V$ which is related to
$N_0$. It is also observed that $V$ is approximately unchanged or
increases slightly with the increase of $|\eta|$ in the considered
$|\eta|$ range. Both $N_0$ and $V$ are related to the data due to
the constraint of normalization.

The values of $N_0$ ($V$) extracted from Eq.~(\ref{eq3}) [Eq.
(\ref{eq5})] using the three-component function (the $q$-dual
function) show some fluctuations at some $|\eta|$ values. This is
caused by the data itself because the two parameters are only the
normalization constant. The parameter values extracted from the
simulation results of the MC models show the same behavior but
generally the values of $T_B$ ($T_q$) for HIJING model are lower
while those of Pythia8 and QGSJETII-04 are higher than that of
$T_B$ ($T_q$) extracted from the data. Furthermore, the values of
$N_0$ ($V$) obtained from the Pythia8 model are very close to the
experimental data whereas HIJING and QGSJETII-04 have lower values
of $N_0$ ($V$) than the data.

%\begin{sidewaystable}
\begin{table*}[]
{\small Table 1. Values of free parameter $T_B$, normalization
constant $N_0$, $\chi^2$ and ndof extracted from the $p_{\rm T}$
spectra of charged hadrons produced in $pp$ collisions at
$\sqrt{s}$~=0.9 TeV using the three-component function.

\begin{center}
\begin{tabular} {cccccccccccc}\\ \hline\hline
Case & $\eta$ & $T_B$ (GeV) & $N_0$ & $\chi^2$/ndof \\
\hline
$         $  & $0.1         $  & $0.163\pm0.001$ & $(7.016\pm0.027)\times10^{-1}$  & $1/8$\\
$         $  & $0.3         $  & $0.164\pm0.002$ & $(7.156\pm0.067)\times10^{-1}$  & $1/8$\\
$         $  & $0.5         $  & $0.164\pm0.001$ & $(7.136\pm0.049)\times10^{-1}$  & $1/8$\\
$         $  & $0.7         $  & $0.163\pm0.001$ & $(7.095\pm0.051)\times10^{-1}$  & $0.4/8$\\
$         $  & $0.9         $  & $0.160\pm0.001$ & $(7.217\pm0.027)\times10^{-1}$  & $0.5/8$\\
Data         & $1.1         $  & $0.159\pm0.001$ & $(7.154\pm0.045)\times10^{-1}$  & $0.4/8$\\
$         $  & $1.3         $  & $0.154\pm0.001$ & $(7.397\pm0.045)\times10^{-1}$  & $0.3/8$\\
$         $  & $1.5         $  & $0.156\pm0.001$ & $(7.415\pm0.039)\times10^{-1}$  & $0.6/8$\\
$         $  & $1.7         $  & $0.151\pm0.001$ & $(7.497\pm0.059)\times10^{-1}$  & $1/8$\\
$         $  & $1.9         $  & $0.152\pm0.001$ & $(7.477\pm0.073)\times10^{-1}$  & $1/8$\\
$         $  & $2.1         $  & $0.149\pm0.001$ & $(7.438\pm0.071)\times10^{-1}$  & $1/8$\\
$         $  & $2.3         $  & $0.148\pm0.002$ & $(7.218\pm0.075)\times10^{-1}$  & $3/8$\\
$         $  & $|\eta| <2.4 $  & $0.158\pm0.002$ & $(1.724\pm0.030)\times10^{1}$   & $14/18$\\
\hline
$         $  & $0.1         $  & $0.159\pm0.002$ & $(6.275\pm0.043)\times10^{-1}$  & $1/8$\\
$         $  & $0.3         $  & $0.160\pm0.001$ & $(6.334\pm0.061)\times10^{-1}$  & $2/8$\\
$         $  & $0.5         $  & $0.157\pm0.001$ & $(6.416\pm0.059)\times10^{-1}$  & $2/8$\\
$         $  & $0.7         $  & $0.156\pm0.001$ & $(6.516\pm0.079)\times10^{-1}$  & $2/8$\\
$         $  & $0.9         $  & $0.154\pm0.001$ & $(6.616\pm0.061)\times10^{-1}$  & $2/8$\\
HIJING       & $1.1         $  & $0.153\pm0.001$ & $(6.716\pm0.039)\times10^{-1}$  & $1/8$\\
$         $  & $1.3         $  & $0.152\pm0.001$ & $(6.776\pm0.039)\times10^{-1}$  & $1/8$\\
$         $  & $1.5         $  & $0.150\pm0.001$ & $(6.837\pm0.035)\times10^{-1}$  & $2/8$\\
$         $  & $1.7         $  & $0.150\pm0.001$ & $(6.877\pm0.039)\times10^{-1}$  & $3/8$\\
$         $  & $1.9         $  & $0.148\pm0.001$ & $(6.897\pm0.025)\times10^{-1}$  & $3/8$\\
$         $  & $2.1         $  & $0.147\pm0.001$ & $(6.838\pm0.027)\times10^{-1}$  & $2/8$\\
$         $  & $2.3         $  & $0.147\pm0.001$ & $(6.777\pm0.055)\times10^{-1}$  & $2/8$\\
$         $  & $|\eta| <2.4 $  & $0.151\pm0.001$ & $(1.591\pm0.027)\times10^{1}$   & $16/18$\\
\hline
$         $  & $0.1         $  & $0.168\pm0.001$ & $(7.017\pm0.081)\times10^{-1}$  & $4/8$\\
$         $  & $0.3         $  & $0.167\pm0.001$ & $(7.038\pm0.055)\times10^{-1}$  & $1/8$\\
$         $  & $0.5         $  & $0.166\pm0.001$ & $(7.118\pm0.039)\times10^{-1}$  & $2/8$\\
$         $  & $0.7         $  & $0.165\pm0.001$ & $(7.198\pm0.063)\times10^{-1}$  & $2/8$\\
$         $  & $0.9         $  & $0.163\pm0.001$ & $(7.378\pm0.047)\times10^{-1}$  & $1/8$\\
Pythia8      & $1.1         $  & $0.162\pm0.001$ & $(7.438\pm0.045)\times10^{-1}$  & $2/8$\\
$         $  & $1.3         $  & $0.161\pm0.001$ & $(7.578\pm0.035)\times10^{-1}$  & $1/8$\\
$         $  & $1.5         $  & $0.159\pm0.001$ & $(7.578\pm0.025)\times10^{-1}$  & $1/8$\\
$         $  & $1.7         $  & $0.156\pm0.001$ & $(7.578\pm0.047)\times10^{-1}$  & $2/8$\\
$         $  & $1.9         $  & $0.155\pm0.001$ & $(7.558\pm0.045)\times10^{-1}$  & $2/8$\\
$         $  & $2.1         $  & $0.154\pm0.001$ & $(7.478\pm0.053)\times10^{-1}$  & $3/8$\\
$         $  & $2.3         $  & $0.153\pm0.001$ & $(7.378\pm0.053)\times10^{-1}$  & $2/8$\\
$         $  & $|\eta| <2.4 $  & $0.161\pm0.001$ & $(1.775\pm0.024)\times10^{1}$   & $7/18$\\
\hline
$           $  & $0.1         $  & $0.176\pm0.001$ & $(5.476\pm0.002)\times10^{-1}$  & $0.4/8$\\
$           $  & $0.3         $  & $0.175\pm0.001$ & $(5.496\pm0.004)\times10^{-1}$  & $0.3/8$\\
$           $  & $0.5         $  & $0.175\pm0.001$ & $(5.575\pm0.004)\times10^{-1}$  & $0.4/8$\\
$           $  & $0.7         $  & $0.171\pm0.001$ & $(5.656\pm0.007)\times10^{-1}$  & $0.1/8$\\
$           $  & $0.9         $  & $0.170\pm0.001$ & $(5.736\pm0.015)\times10^{-1}$  & $0.4/8$\\
QGSJETII-04    & $1.1         $  & $0.168\pm0.001$ & $(5.797\pm0.019)\times10^{-1}$  & $0.6/8$\\
$           $  & $1.3         $  & $0.167\pm0.001$ & $(5.875\pm0.023)\times10^{-1}$  & $1/8$\\
$           $  & $1.5         $  & $0.165\pm0.001$ & $(5.895\pm0.019)\times10^{-1}$  & $1/8$\\
$           $  & $1.7         $  & $0.164\pm0.001$ & $(5.915\pm0.029)\times10^{-1}$  & $1/8$\\
$           $  & $1.9         $  & $0.164\pm0.001$ & $(5.891\pm0.019)\times10^{-1}$  & $1/8$\\
$           $  & $2.1         $  & $0.166\pm0.001$ & $(5.863\pm0.025)\times10^{-1}$  & $1/8$\\
$           $  & $2.3         $  & $0.165\pm0.001$ & $(5.784\pm0.059)\times10^{-1}$  & $2/8$\\
$           $  & $|\eta| <2.4 $  & $0.167\pm0.001$ & $(1.386\pm0.005)\times10^{1}$   & $4/18$\\
\hline

\end{tabular}%
\end{center}}
\end{table*}
%\end{sidewaystable}

%\begin{sidewaystable}
\begin{table*}[]
{\small Table 2. Values of free parameters $T_q$ and $q$, kinetic
freeze-out volume $V$, $\chi^2$ and ndof extracted from the
$p_{\rm T}$ spectra of charged hadrons produced in $pp$ collisions
at $\sqrt{s}$~=0.9 TeV using the $q$-dual function.

\begin{center}
\begin{tabular} {cccccccccccc}\\ \hline\hline
Case & $\eta$ & $T_q$ (GeV) & $q$  & $V$ (fm$^3$) & $\chi^2$/ndof \\
\hline
$         $  & $0.1         $  & $0.100\pm0.001$ & $1.134\pm0.001$  &$(4.339\pm0.037)\times10^{4}$  & $1/11$\\
$         $  & $0.3         $  & $0.100\pm0.001$ & $1.135\pm0.001$  &$(4.353\pm0.039)\times10^{4}$  & $1/11$\\
$         $  & $0.5         $  & $0.100\pm0.001$ & $1.135\pm0.001$  &$(4.390\pm0.043)\times10^{4}$  & $2/11$\\
$         $  & $0.7         $  & $0.098\pm0.001$ & $1.135\pm0.002$  &$(4.685\pm0.044)\times10^{4}$  & $1/11$\\
$         $  & $0.9         $  & $0.098\pm0.001$ & $1.134\pm0.001$  &$(4.787\pm0.046)\times10^{4}$  & $1/11$\\
Data         & $1.1         $  & $0.097\pm0.001$ & $1.134\pm0.001$  &$(4.837\pm0.072)\times10^{4}$  & $2/11$\\
$         $  & $1.3         $  & $0.096\pm0.001$ & $1.133\pm0.001$  &$(5.206\pm0.070)\times10^{4}$  & $1/11$\\
$         $  & $1.5         $  & $0.095\pm0.001$ & $1.133\pm0.001$  &$(5.417\pm0.069)\times10^{4}$  & $1/11$\\
$         $  & $1.7         $  & $0.094\pm0.001$ & $1.133\pm0.001$  &$(5.647\pm0.071)\times10^{4}$  & $2/11$\\
$         $  & $1.9         $  & $0.093\pm0.001$ & $1.133\pm0.001$  &$(5.872\pm0.072)\times10^{4}$  & $1/11$\\
$         $  & $2.1         $  & $0.092\pm0.001$ & $1.132\pm0.002$  &$(6.169\pm0.082)\times10^{4}$  & $2/11$\\
$         $  & $2.3         $  & $0.091\pm0.001$ & $1.132\pm0.002$  &$(6.163\pm0.087)\times10^{4}$  & $2/11$\\
$         $  & $|\eta| <2.4 $  & $0.130\pm0.001$ & $1.113\pm0.002$  &$(2.267\pm0.045)\times10^{4}$  & $22/21$\\
\hline
$         $  & $0.1         $  & $0.115\pm0.001$ & $1.096\pm0.002$  &$(2.771\pm0.098)\times10^{4}$  & $44/11$\\
$         $  & $0.3         $  & $0.115\pm0.001$ & $1.096\pm0.003$  &$(2.790\pm0.113)\times10^{4}$  & $55/11$\\
$         $  & $0.5         $  & $0.115\pm0.001$ & $1.096\pm0.002$  &$(2.810\pm0.099)\times10^{4}$  & $58/11$\\
$         $  & $0.7         $  & $0.115\pm0.001$ & $1.093\pm0.003$  &$(2.907\pm0.110)\times10^{4}$  & $61/11$\\
$         $  & $0.9         $  & $0.115\pm0.001$ & $1.093\pm0.003$  &$(2.936\pm0.113)\times10^{4}$  & $58/11$\\
HIJING       & $1.1         $  & $0.113\pm0.001$ & $1.094\pm0.003$  &$(3.160\pm0.102)\times10^{4}$  & $61/11$\\
$         $  & $1.3         $  & $0.112\pm0.001$ & $1.094\pm0.003$  &$(3.324\pm0.099)\times10^{4}$  & $53/11$\\
$         $  & $1.5         $  & $0.110\pm0.001$ & $1.095\pm0.002$  &$(3.606\pm0.144)\times10^{4}$  & $45/11$\\
$         $  & $1.7         $  & $0.110\pm0.001$ & $1.095\pm0.002$  &$(3.592\pm0.099)\times10^{4}$  & $45/11$\\
$         $  & $1.9         $  & $0.110\pm0.001$ & $1.097\pm0.002$  &$(3.530\pm0.121)\times10^{4}$  & $38/11$\\
$         $  & $2.1         $  & $0.110\pm0.001$ & $1.095\pm0.002$  &$(3.557\pm0.101)\times10^{4}$  & $31/11$\\
$         $  & $2.3         $  & $0.110\pm0.001$ & $1.096\pm0.003$  &$(3.458\pm0.165)\times10^{4}$  & $46/11$\\
$         $  & $|\eta| <2.4 $  & $0.113\pm0.001$ & $1.123\pm0.001$  &$(2.110\pm0.063)\times10^{4}$  & $51/21$\\
\hline
$         $  & $0.1         $  & $0.122\pm0.001$ & $1.093\pm0.003$  &$(2.715\pm0.094)\times10^{4}$  & $27/11$\\
$         $  & $0.3         $  & $0.122\pm0.001$ & $1.093\pm0.002$  &$(2.732\pm0.096)\times10^{4}$  & $23/11$\\
$         $  & $0.5         $  & $0.122\pm0.001$ & $1.090\pm0.003$  &$(2.796\pm0.072)\times10^{4}$  & $21/11$\\
$         $  & $0.7         $  & $0.122\pm0.001$ & $1.089\pm0.001$  &$(2.836\pm0.071)\times10^{4}$  & $19/11$\\
$         $  & $0.9         $  & $0.122\pm0.001$ & $1.088\pm0.001$  &$(2.925\pm0.091)\times10^{4}$  & $18/11$\\
Pythia8      & $1.1         $  & $0.122\pm0.001$ & $1.087\pm0.002$  &$(2.931\pm0.092)\times10^{4}$  & $22/11$\\
$         $  & $1.3         $  & $0.122\pm0.001$ & $1.086\pm0.003$  &$(3.036\pm0.134)\times10^{4}$  & $16/11$\\
$         $  & $1.5         $  & $0.121\pm0.001$ & $1.085\pm0.003$  &$(3.086\pm0.074)\times10^{4}$  & $20/11$\\
$         $  & $1.7         $  & $0.121\pm0.001$ & $1.082\pm0.003$  &$(3.145\pm0.071)\times10^{4}$  & $15/11$\\
$         $  & $1.9         $  & $0.121\pm0.001$ & $1.081\pm0.002$  &$(3.141\pm0.072)\times10^{4}$  & $19/11$\\
$         $  & $2.1         $  & $0.121\pm0.001$ & $1.080\pm0.003$  &$(3.102\pm0.088)\times10^{4}$  & $24/11$\\
$         $  & $2.3         $  & $0.120\pm0.001$ & $1.079\pm0.002$  &$(3.172\pm0.085)\times10^{4}$  & $19/11$\\
$         $  & $|\eta| <2.4 $  & $0.130\pm0.001$ & $1.107\pm0.001$  &$(2.216\pm0.054)\times10^{4}$  & $61/21$\\
\hline
$           $  & $0.1         $  & $0.140\pm0.001$ & $1.077\pm0.002$  &$(1.406\pm0.031)\times10^{4}$  & $30/11$\\
$           $  & $0.3         $  & $0.140\pm0.001$ & $1.077\pm0.002$  &$(1.405\pm0.030)\times10^{4}$  & $30/11$\\
$           $  & $0.5         $  & $0.140\pm0.001$ & $1.076\pm0.003$  &$(1.442\pm0.061)\times10^{4}$  & $35/11$\\
$           $  & $0.7         $  & $0.137\pm0.001$ & $1.078\pm0.002$  &$(1.561\pm0.040)\times10^{4}$  & $28/11$\\
$           $  & $0.9         $  & $0.137\pm0.001$ & $1.076\pm0.003$  &$(1.593\pm0.044)\times10^{4}$  & $35/11$\\
QGSJETII-04    & $1.1         $  & $0.134\pm0.001$ & $1.080\pm0.002$  &$(1.690\pm0.047)\times10^{4}$  & $30/11$\\
$           $  & $1.3         $  & $0.133\pm0.001$ & $1.080\pm0.002$  &$(1.748\pm0.048)\times10^{4}$  & $33/11$\\
$           $  & $1.5         $  & $0.131\pm0.001$ & $1.081\pm0.002$  &$(1.833\pm0.048)\times10^{4}$  & $31/11$\\
$           $  & $1.7         $  & $0.131\pm0.001$ & $1.079\pm0.002$  &$(1.852\pm0.052)\times10^{4}$  & $36/11$\\
$           $  & $1.9         $  & $0.131\pm0.001$ & $1.077\pm0.003$  &$(1.848\pm0.051)\times10^{4}$  & $42/11$\\
$           $  & $2.1         $  & $0.129\pm0.001$ & $1.078\pm0.002$  &$(1.867\pm0.050)\times10^{4}$  & $43/11$\\
$           $  & $2.3         $  & $0.126\pm0.001$ & $1.080\pm0.002$  &$(2.035\pm0.039)\times10^{4}$  & $35/11$\\
$           $  & $|\eta| <2.4 $  & $0.127\pm0.001$ & $1.120\pm0.002$  &$(1.741\pm0.006)\times10^{4}$  & $112/21$\\
\hline
\end{tabular}%
\end{center}}
\end{table*}
%\end{sidewaystable}

\begin{figure*}[]
\centering
\includegraphics[width=0.32\textwidth]{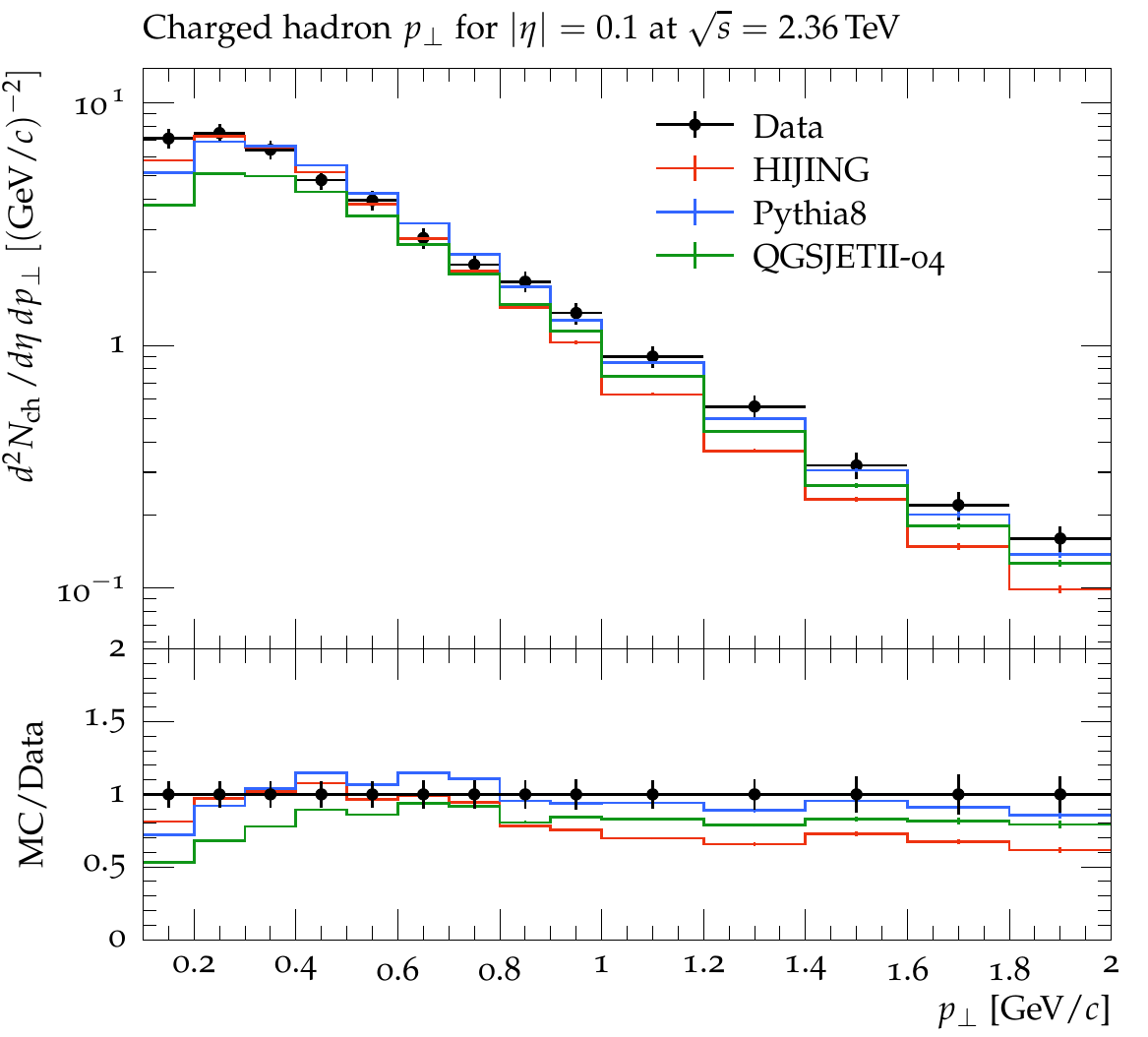}
\includegraphics[width=0.32\textwidth]{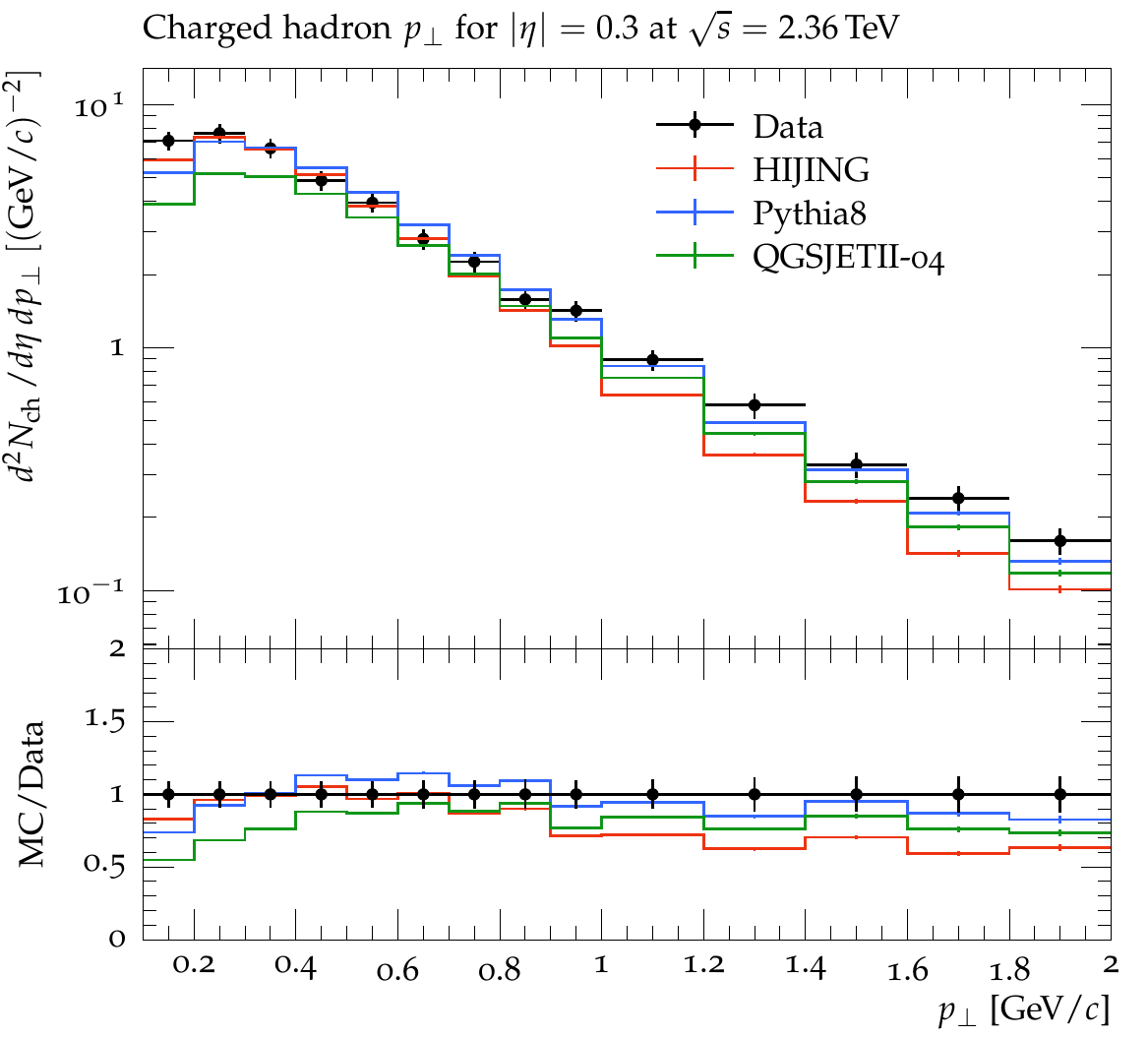}
\includegraphics[width=0.32\textwidth]{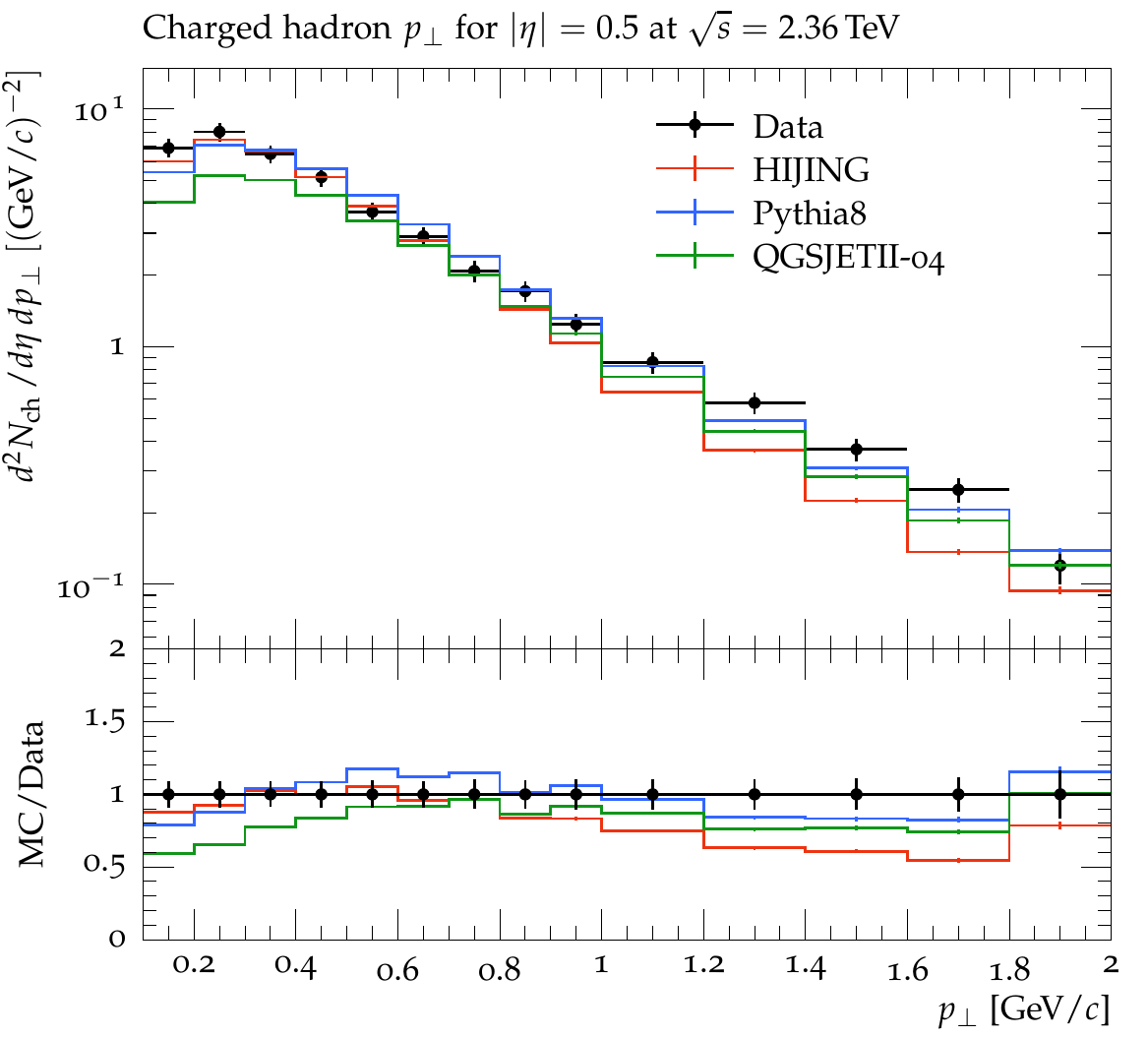}
\includegraphics[width=0.32\textwidth]{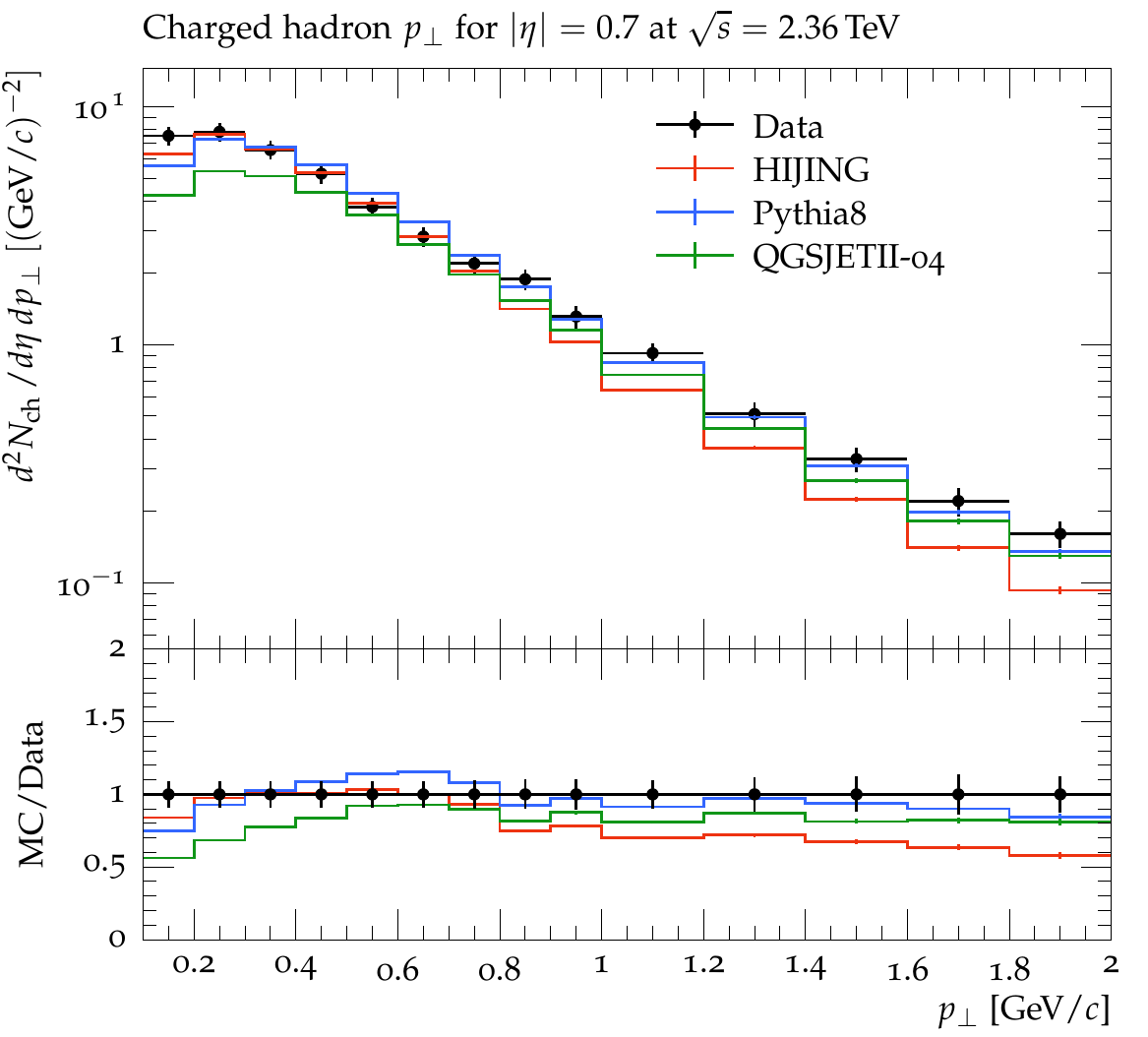}
\includegraphics[width=0.32\textwidth]{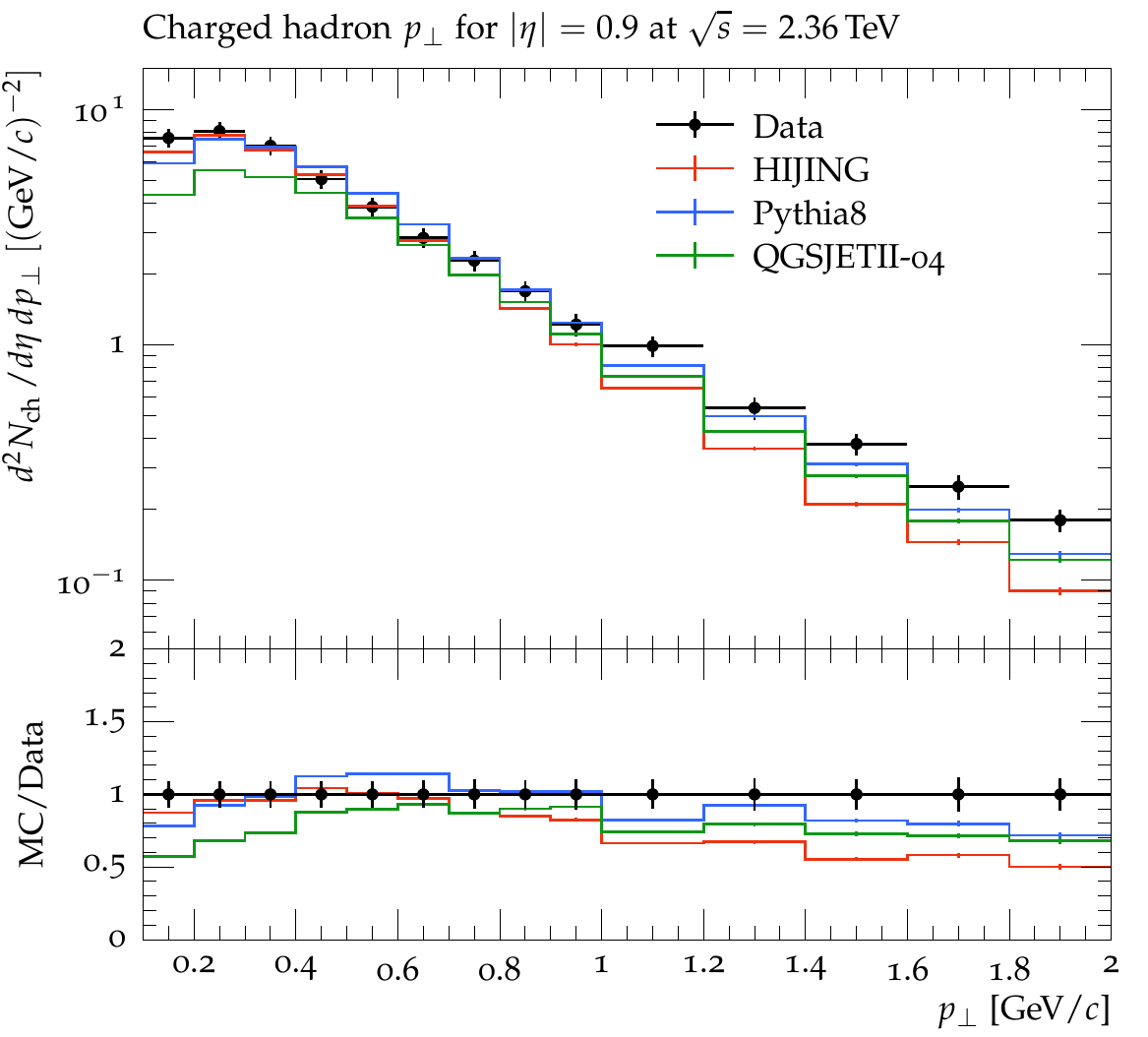}
\includegraphics[width=0.32\textwidth]{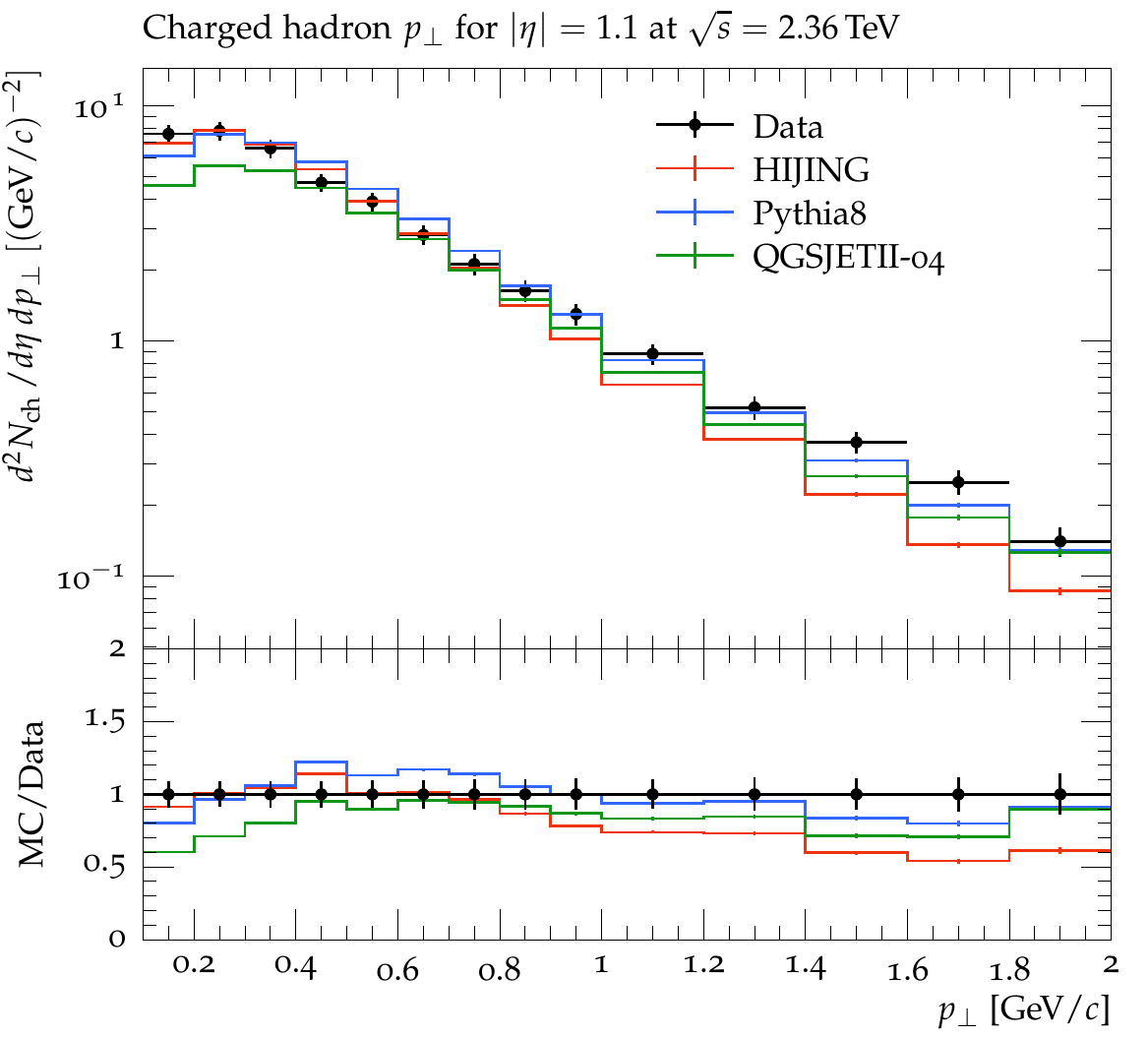}
\includegraphics[width=0.32\textwidth]{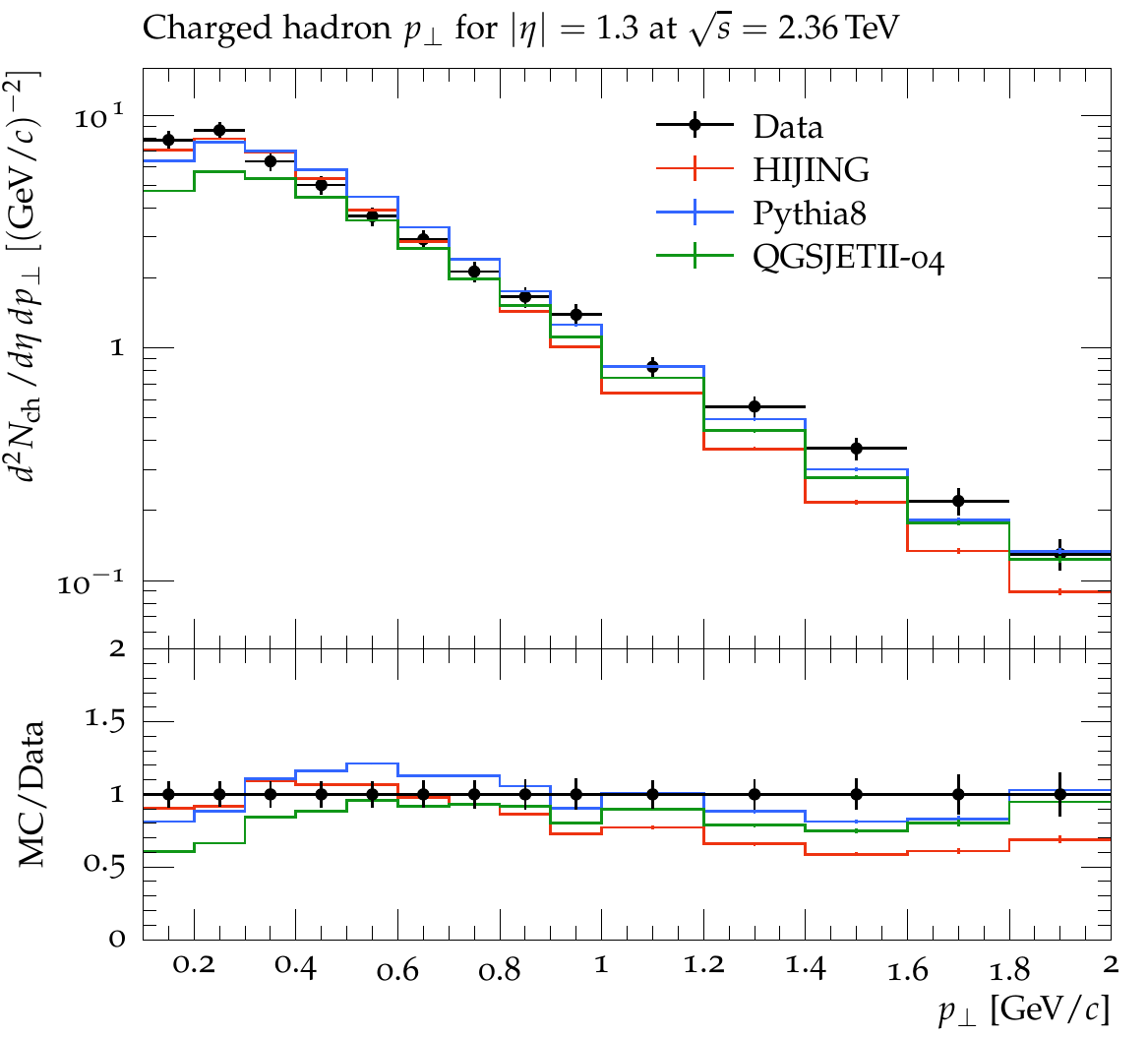}
\includegraphics[width=0.32\textwidth]{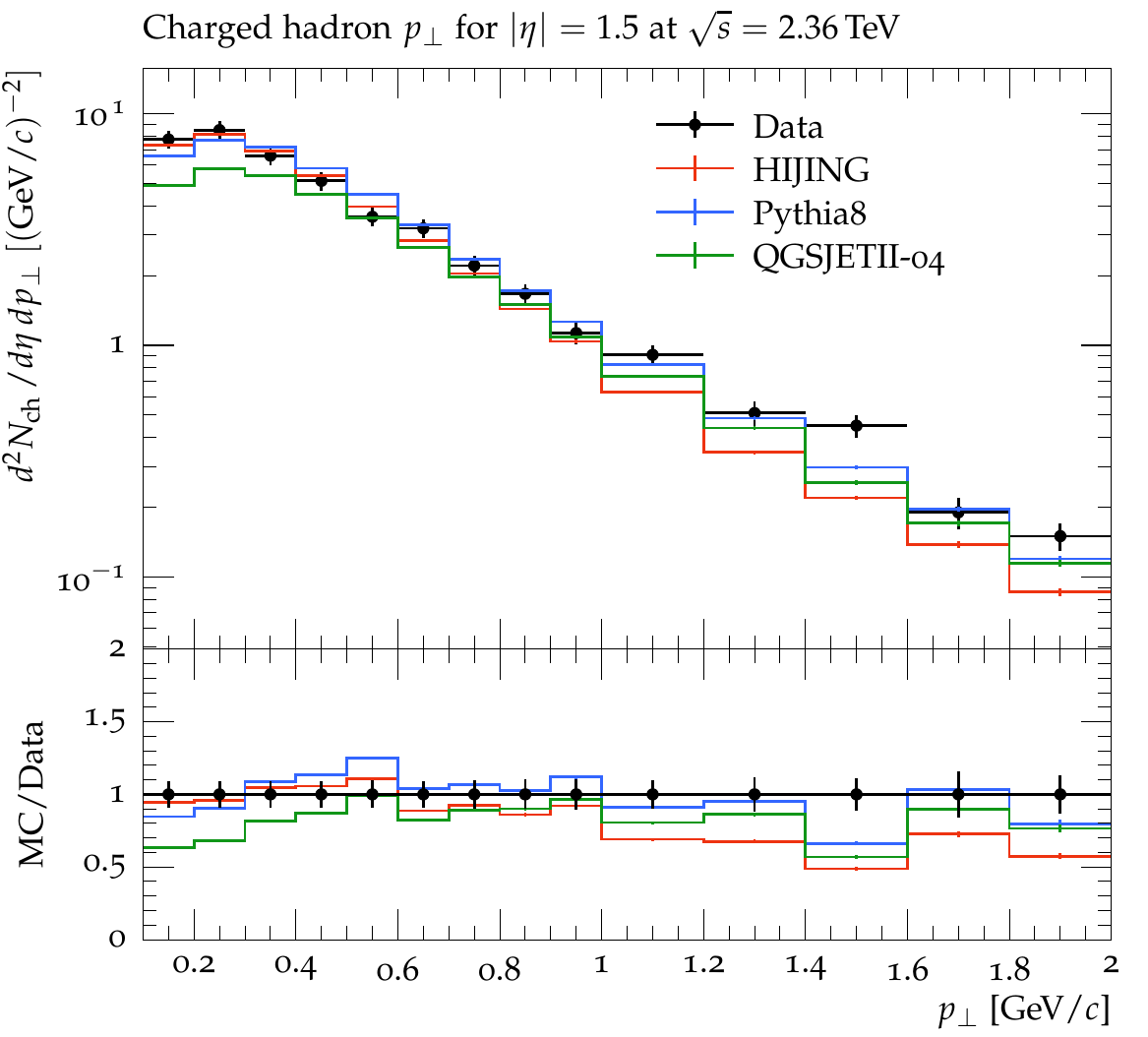}
\includegraphics[width=0.32\textwidth]{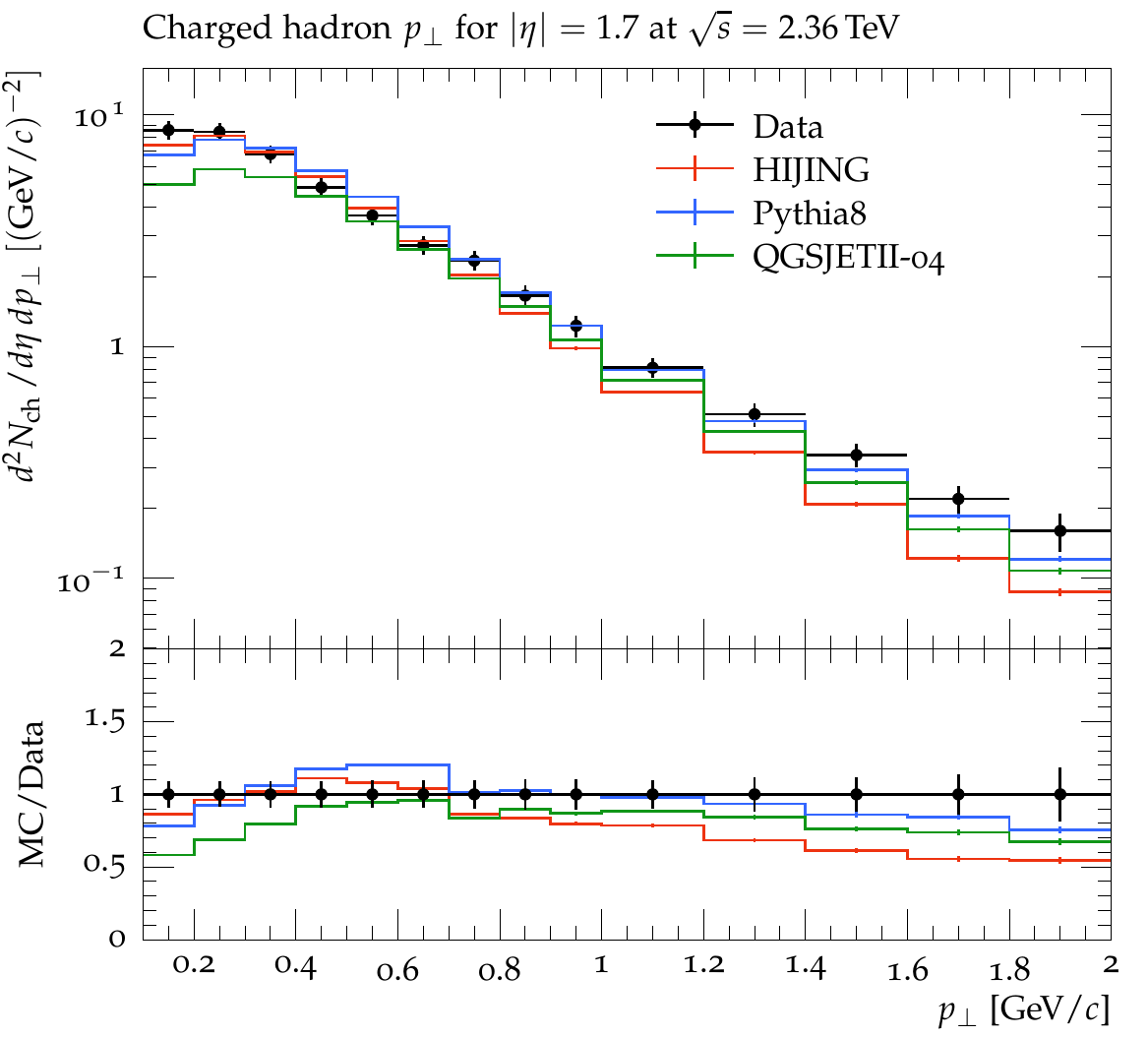}
\includegraphics[width=0.32\textwidth]{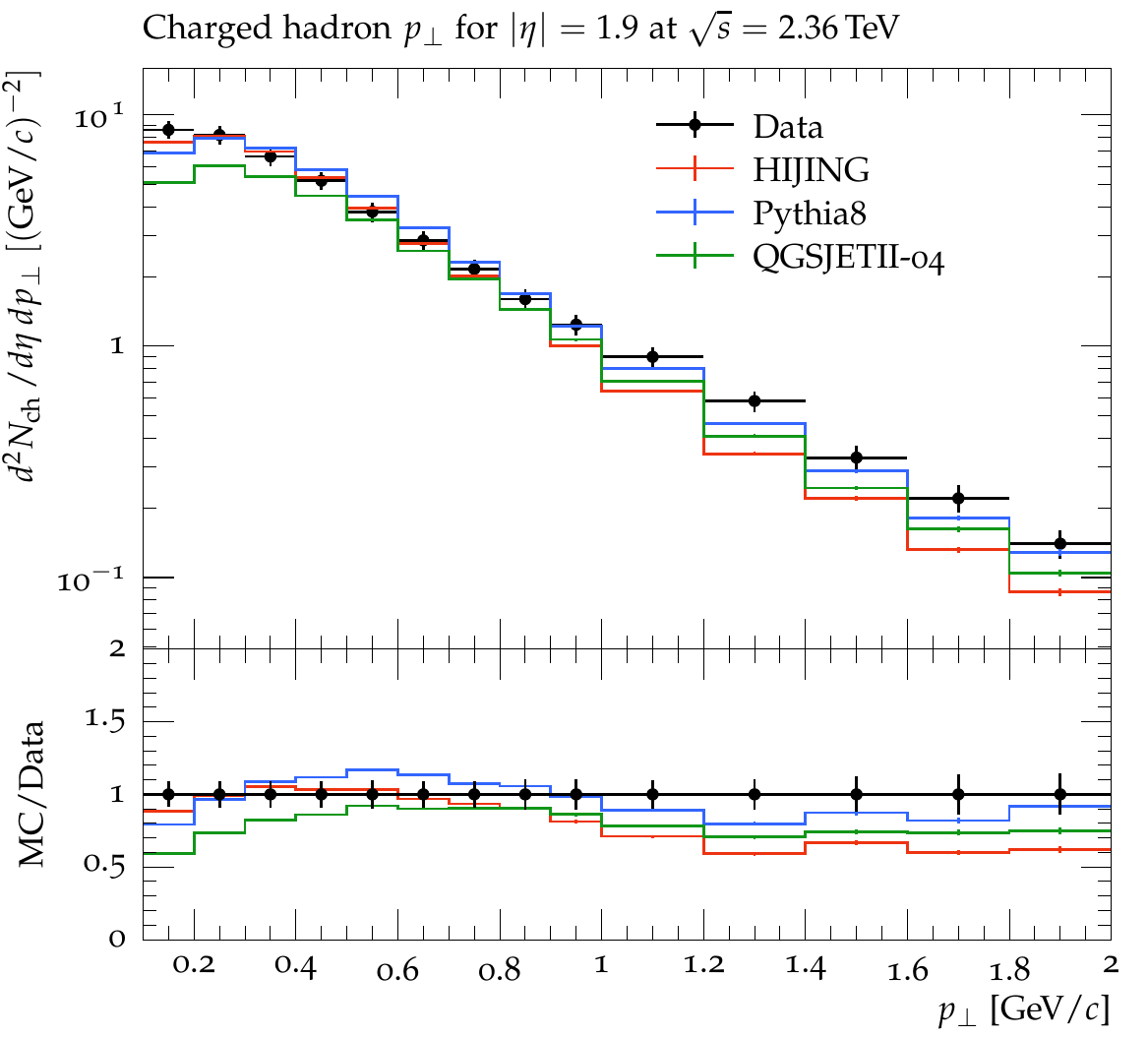}
\includegraphics[width=0.32\textwidth]{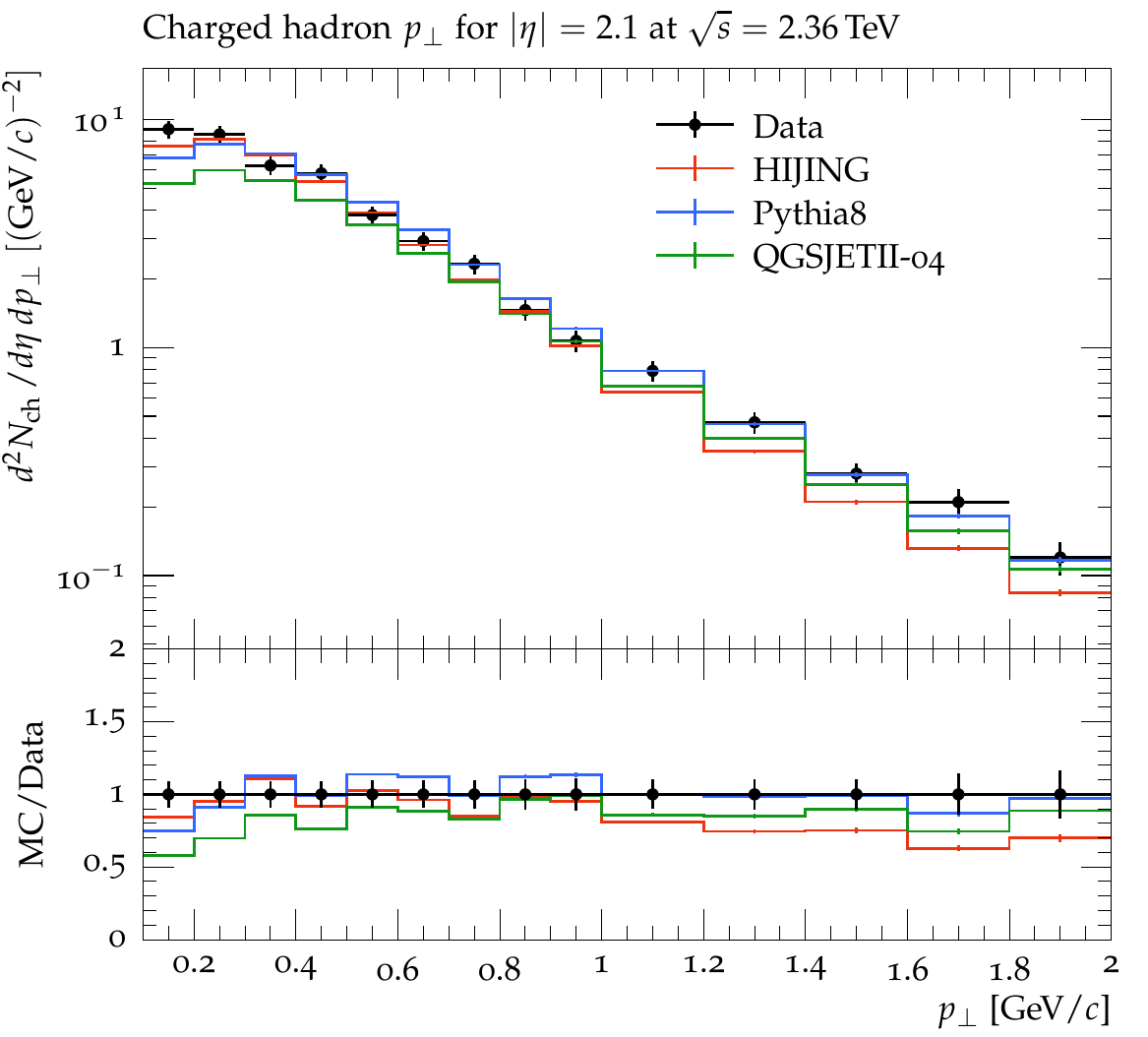}
\includegraphics[width=0.32\textwidth]{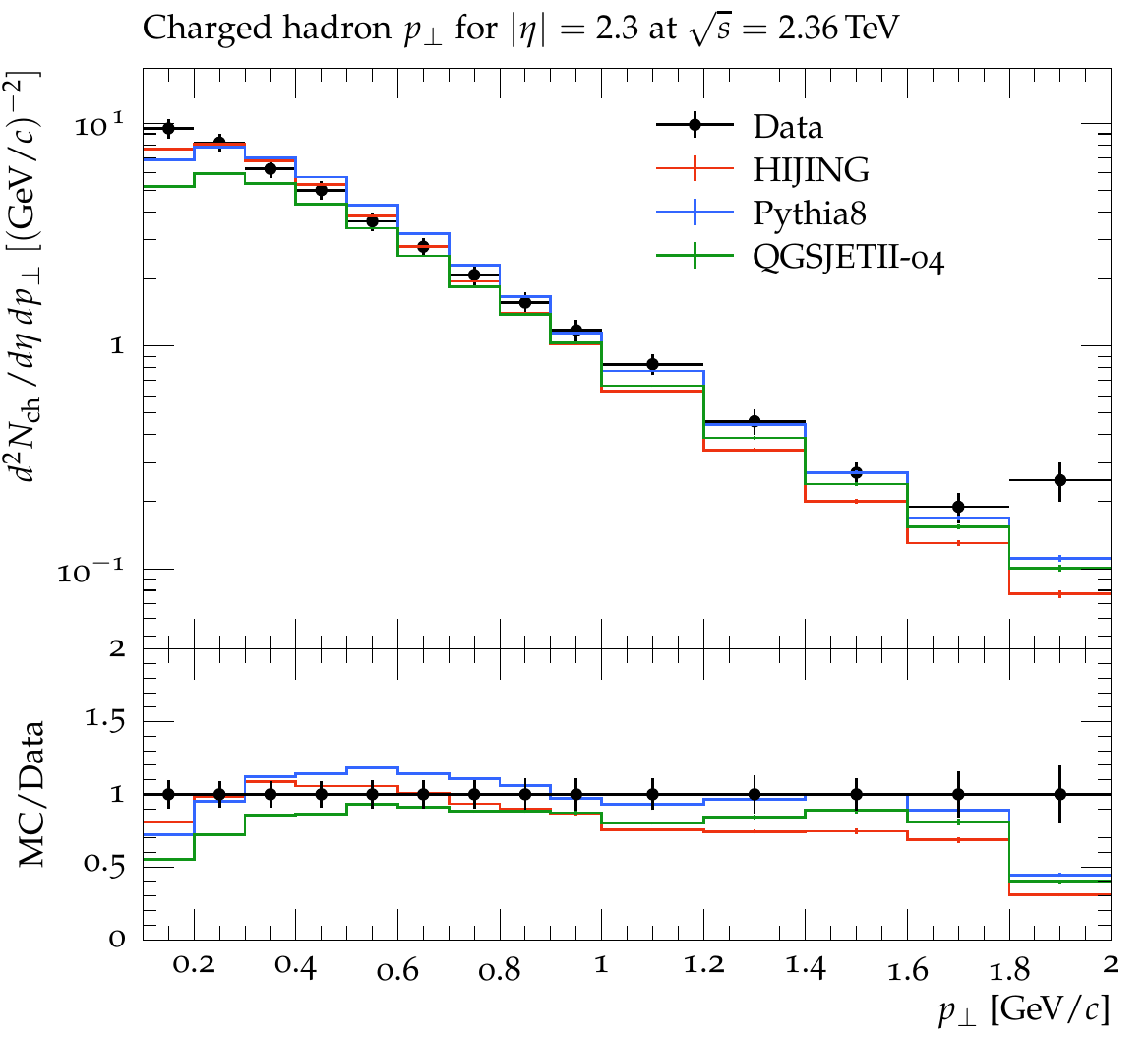}
\caption{The transverse momentum spectra shown in different $|\eta|$ bins of charged hadrons \cite{cms} are compared with the
MC model predictions \cite{hijing,string,qgsjetII1} from $pp$
collisions at $\sqrt{s}$~= 2.36 TeV. The solid black markers show the experimental data while the solid lines of different colors represent the different MC model predictions shown in the panels.} \label{fig3}
\end{figure*}

\begin{figure*}[]
\centering
\includegraphics[width=0.95\textwidth]{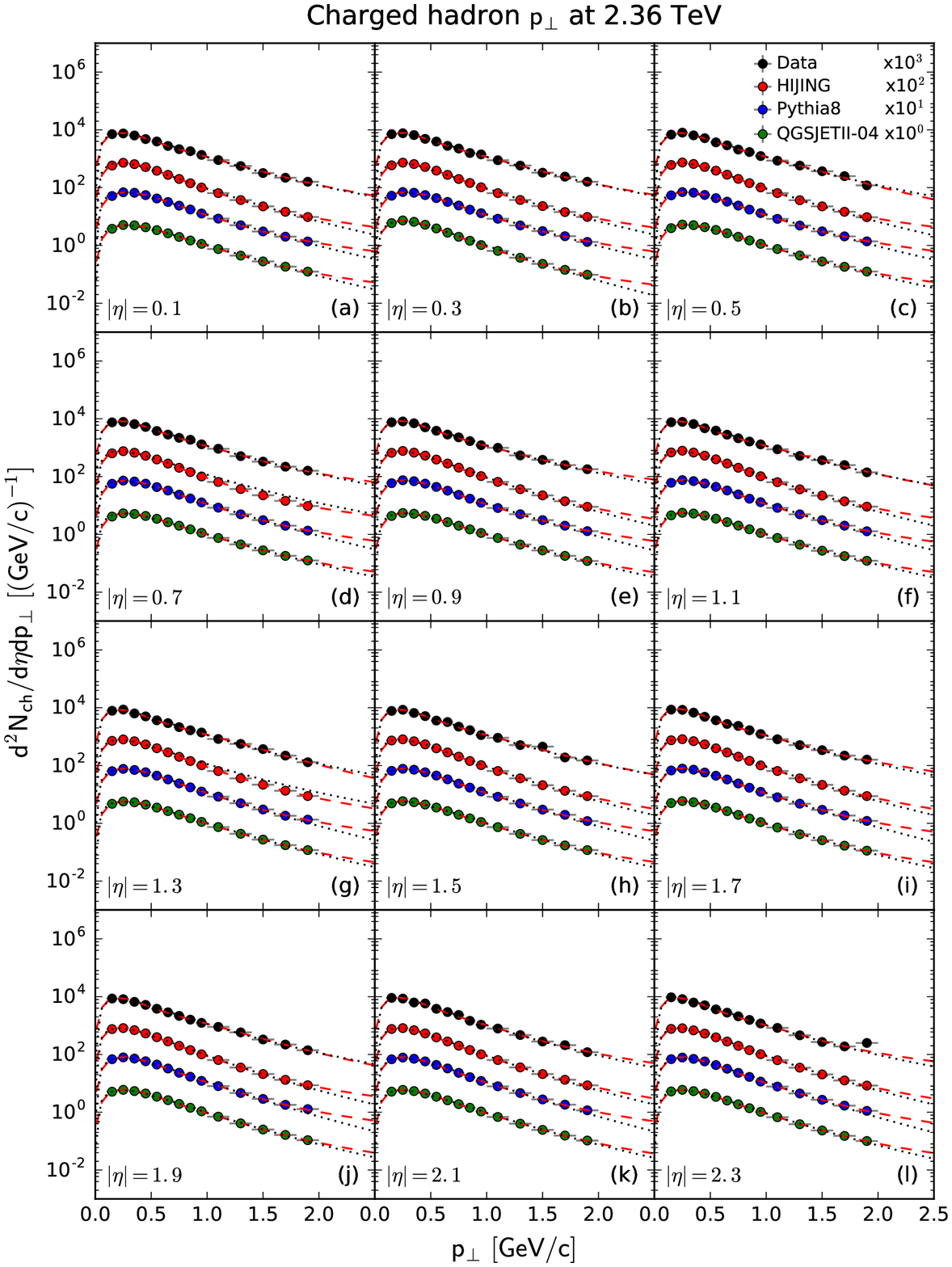}
\caption{The experimental measurements \cite{cms} as well as the
MC model predictions \cite{hijing,string,qgsjetII1} of the
transverse momentum {\ppt} spectra shown in different $|\eta|$
bins of charged hadrons in $pp$ collision at $\sqrt{s}$~= 2.36 TeV are
fitted with two analytic functions. The dashed curves represent
the fit results of the three-component function [Eq. (3)], while
the dotted curves represent the fit results of the $q$-dual
function [Eq. (5)].} \label{fig4}
\end{figure*}

Similar to Figure~\ref{fig1}, Figure~\ref{fig3} shows the {\ppt}
distributions of charged hadrons in several $|\eta|$ bins starting
from 0.1 ($|\eta|=0.0$--0.2) to 2.3 ($|\eta|=2.2$--2.4) in $pp$
collisions at $\sqrt{s}$~= 2.36 TeV. The results corresponding to the
experimental data and the MC model predictions are presented with
the same meaning and style as Figure~\ref{fig1}. The values of
parameters are listed in Table 3. Similar to Figure~\ref{fig2},
Figure~\ref{fig4} shows the {\ppt} distributions of charged
hadrons in several $|\eta|$ bins starting from 0.1 to 2.3 in $pp$
collisions at $\sqrt{s}$~= 2.36 TeV. The results corresponding to the
experimental data, the MC model predictions and the fits by the
two analytic functions are presented with the same meaning and
style as Figure~\ref{fig2}. The values of parameters are listed in
Table 4.

From Figure~\ref{fig3} one can see that the HIJING model
reproduces the measurements for $|\eta|$ from 0.1 to 0.7 between
the {\ppt} from 0.3 to 0.7 GeV/$c$ while up to 0.8 GeV/$c$ for 0.8
$\le$ $\eta$ $\le$ 1.9 and up to 1 GeV/$c$ at $|\eta|=2.1$ and
2.3. This shows that the prediction of the HIJING model gets
better with increasing the value of $|\eta|$ at $\sqrt{s}=2.36$ TeV
as was observed in the case of $\sqrt{s}$~= 0.9 TeV as well. The
prediction of the Pythia model has the closest predictions amongst
the models for the $\sqrt{s}$~= 2.36 TeV energy as well as had better
prediction for the $\sqrt{s}$~= 0.9 TeV. The model depicts the $p_T$
spectra between 0.2--0.4 and 0.8--1.8 GeV/$c$ for $|\eta| = 0.1$
which cover the maximum range of {\ppt}. It overestimates between
0.4--0.8 while sightly underestimates at 0.1 and 2.0 GeV/$c$. The
prediction of the model has little variation in the {\ppt} spectra
with increasing $|\eta|$ and hence has similar behavior for all
$|\eta|$ bins. The model prediction has further improved with
increasing the energy from $\sqrt{s}$~= 0.9 to 2.36 TeV. Finally the
QGSJET model depicts the data for {\ppt} = 0.6--0.8 GeV/$c$ at
$|\eta|$ = 0.1 but underestimates below 0.6 GeV/$c$ and above 0.8
GeV/$c$. Again there is little variation in the behavior of {\ppt}
distribution of the model prediction with varying $|\eta|$. As
with the $\sqrt{s}$~=0.9 TeV data, the prediction of the QGSJET model is
better than the HIJING at low {\ppt} while at high {\ppt}, HIJING
has better results than the QGSJET model. With increasing the
energy, the result of the QGSJET model get worst while HIJING and
Pythia8 models results have been improved.

From Figure~\ref{fig4} one can see that the two analytic functions
fit well the CMS data and the three MC model results. Conclusions
obtained from Figure~\ref{fig2} can be also obtained from
Figure~\ref{fig4}, though the energy-dependent parameters can be
seen from Tables 3 and 4.

From Table 3 one can see that the extracted $T_B$ initially
increases for $|\eta|$ up to 0.9 and then decreases monotonically.
The values of $N_0$ extracted from Eq. 3 using three-component
function in some cases increases with increasing $|\eta|$. The
parameter values extracted from the simulation results of the
models show the same behavior but generally, the values of $T_B$
for the HIJING model are lower while that of the Pythia8 and QGSJETII-04
are higher than that extracted from the data. Furthermore, the
values of $N_0$ obtained from the Pythia8 model are very close to
the experimental data whereas HIJING and QGSJETII-04 have lower
values of $N_0$ than the data.

From Table 4 one can see the values of the parameters, $T_q$, $q$
and $V$, extracted from the data as well as the MC models with the
$q$-dual function [Eq. (5)] at different values of $|\eta|$ at
$\sqrt{s}$ = 2.36 TeV. The value of $T_q$ extracted from the
experimental data decreases monotonically with increasing
$|\eta|$. In the case of models, $T_q$ has also decreasing behavior
with $|\eta|$. The values of $T_q$ for the HIJING model are closer to
the data than the other two models, where QGSJETII-04 has the
highest departure. As before, the value of $q$ has little effect
on varying $|\eta|$ and has therefore almost the same value at
different values of $|\eta|$. The value of $q$ extracted from the
data is higher than those from the MC models. From the models, $q$
has comparable value with Pythia having higher than QGSJETII-04
and lower than the HIJING model. The same has been observed at
$\sqrt{s}$~= 0.9 TeV as well. The kinetic freeze-out volume $V$ is an
increasing function of the $|\eta|$. The increasing behavior of
$V$ extracted by the $q$-dual function is reflected in the data as
well as in the MC models. Generally, the data values are larger
compared to the models among which Pythia8 has higher values of
$V$ than QGSJETII-04 while lower values than HIJING. Although
there is a change in the absolute values of the parameters, the
behavior remained the same at $\sqrt{s}$~= 2.36 TeV as was observed in
the case of $\sqrt{s}$~= 0.9 TeV.

\begin{figure*}[!ht]
\centering
\includegraphics[width=0.49\textwidth]{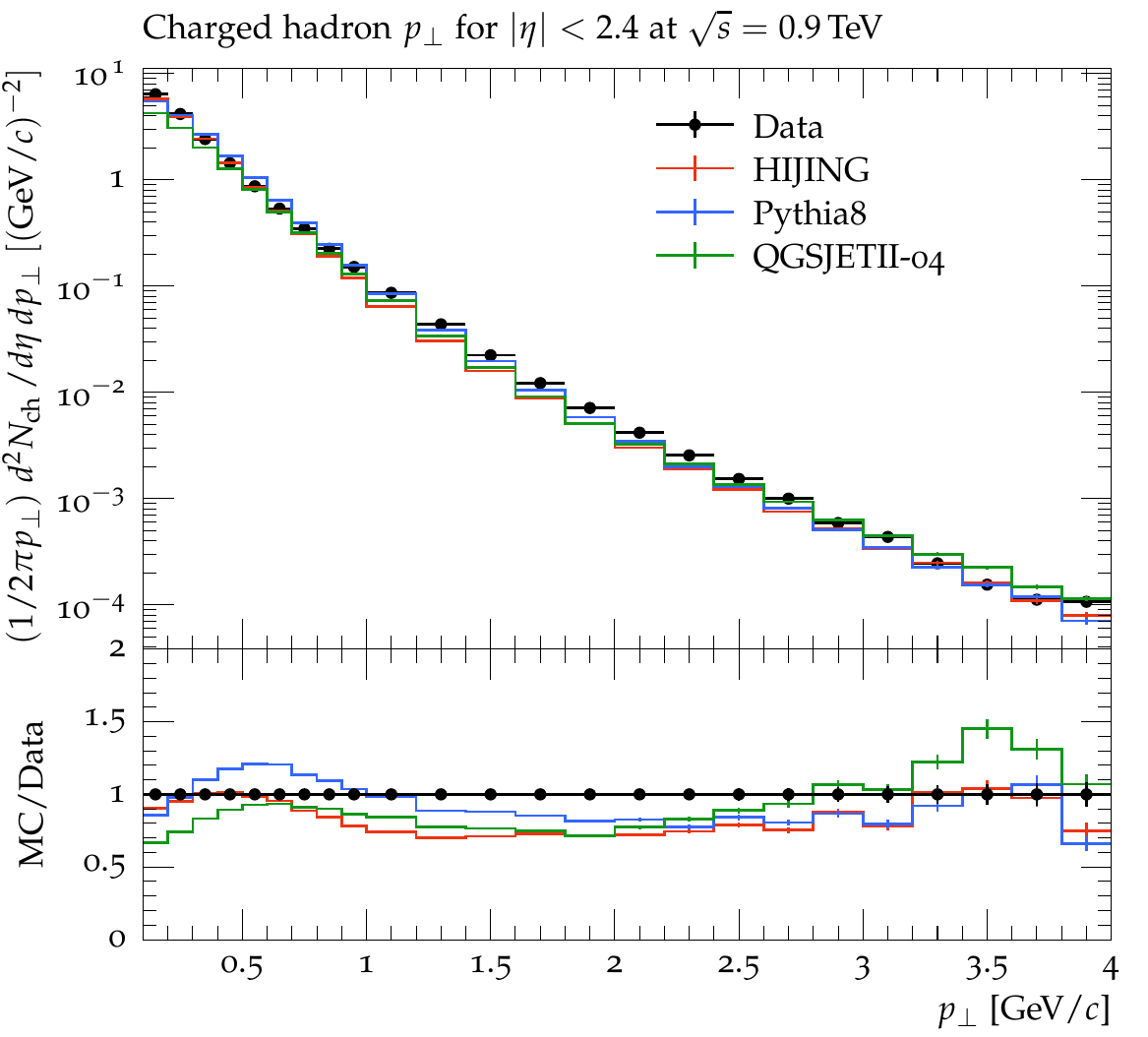}
\includegraphics[width=0.49\textwidth]{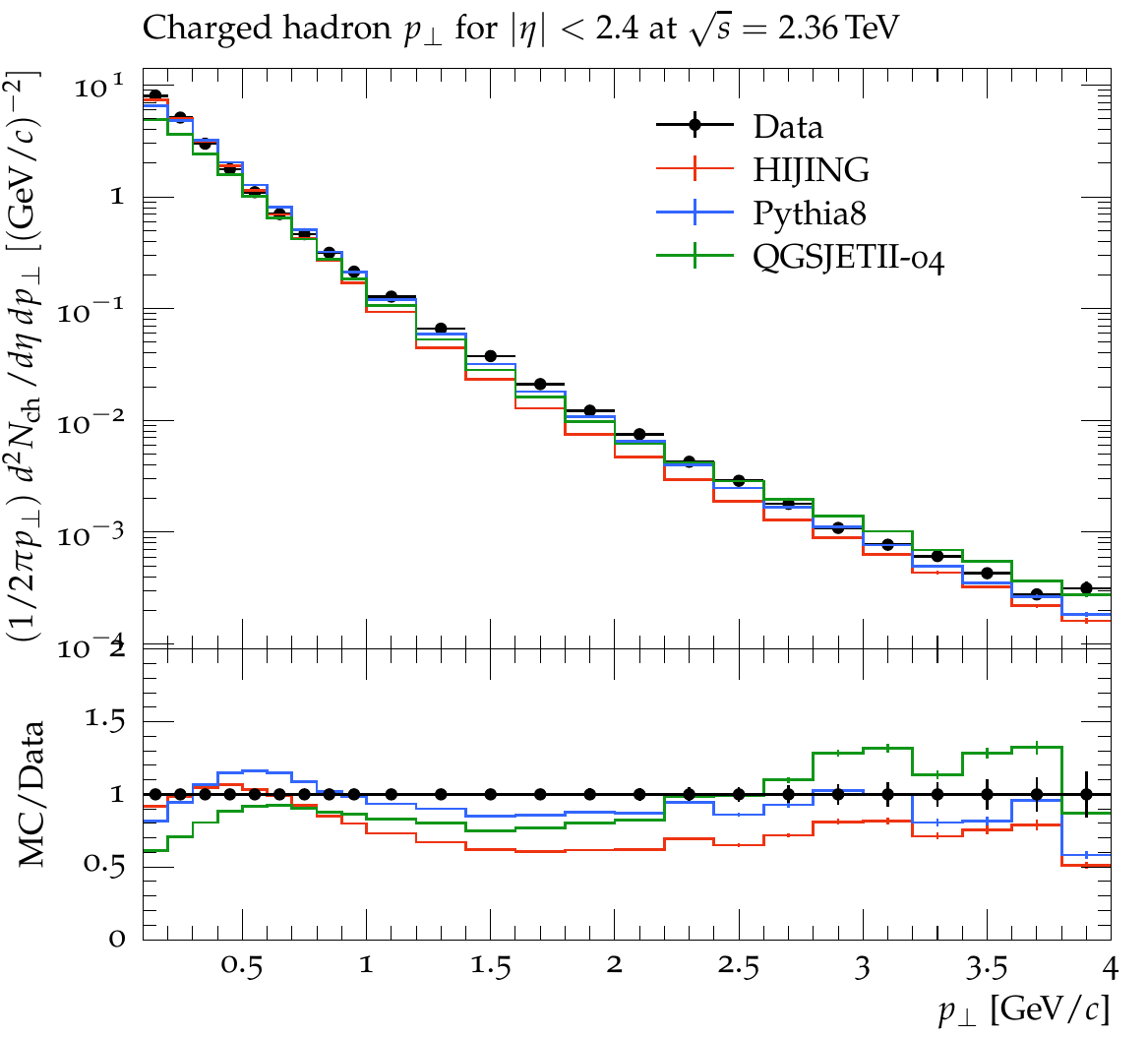}
\caption{The experimental measurements as well as the MC model
predictions of the transverse momentum spectra shown in one wider
$|\eta|$ bin ($0<|\eta|<2.4$) of charged hadrons in $pp$
collisions at $\sqrt{s}$~= 0.9 (left) and 2.36 TeV (right).}
\label{fig5}
\end{figure*}

Figure \ref{fig5} shows the transverse momentum $p_{\rm T}$
spectra of charged hadrons in $pp$ collisions at $\sqrt{s}$~= 0.9
(left) and 2.36 TeV (right) in full $\eta$ bin, $|\eta|<2.4$ with an enhanced $p_T$ range of (0.1 - 4.0) GeV/c. The
measurement shown by the black filled circles is compared with the
MC model predictions that are shown by color lines. The lower panel in
each case is the ratio of the MC predictions to the data
measurements. Pythia8 has a better comparison with data than the
other two models and the prediction has further been slightly
improved with the increase in energy from $\sqrt{s}$~= 0.9 to 2.36 TeV.
After Pythia, HIJING reproduced better comparison which has also
been improved with the increase in energy. Like other models, the
QGSJET model has also reproduced comparable predictions for the
data at $\sqrt{s}$~= 0.9 TeV but at the 2.36 TeV, the prediction for
$p_T$ above 1 GeV/$c$ got worst and are therefore the deviation
from the measurements is more pronounced.

\begin{figure*}[!ht]
%\vskip.2cm
\centering
\includegraphics[width=0.95\textwidth]{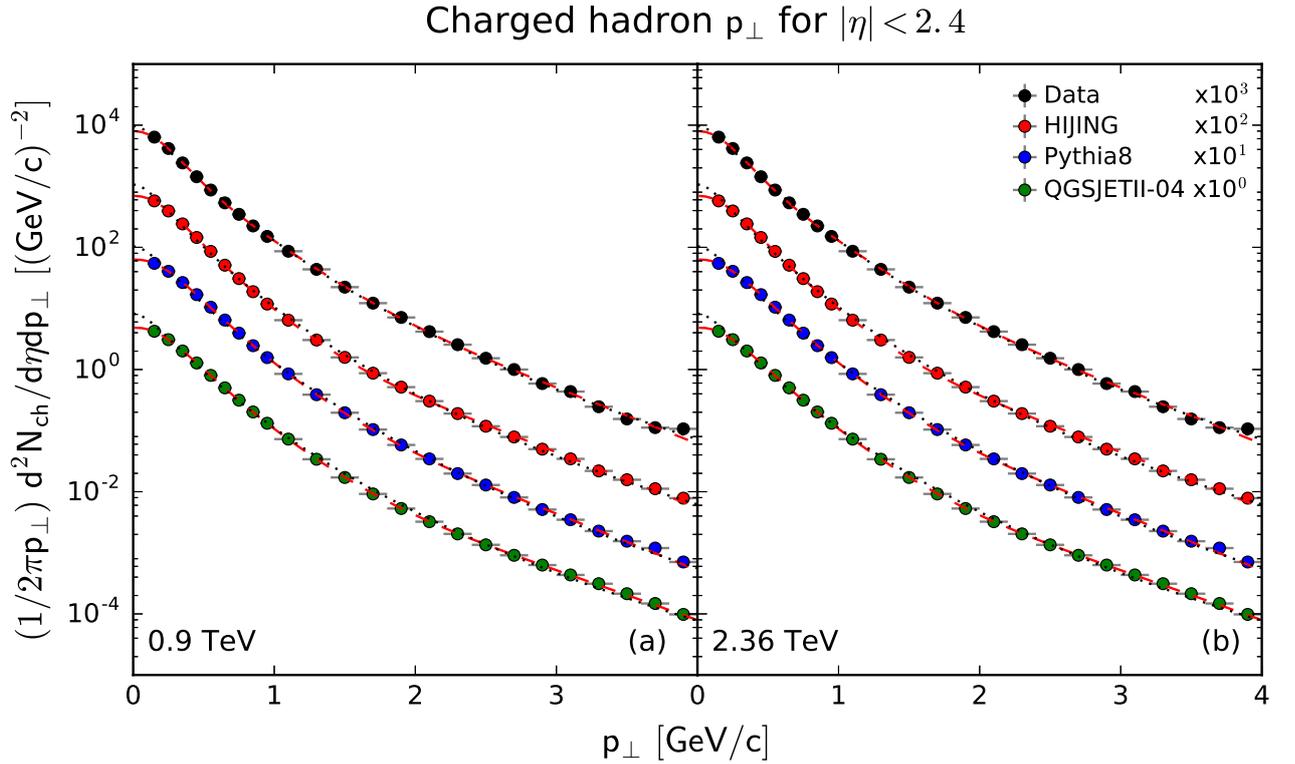}
\caption{The experimental measurements as well as the MC model
predictions of the transverse momentum spectra shown in one wider
$|\eta|$ bin ($0<|\eta|<2.4$) of charged hadrons in $pp$
collisions at $\sqrt{s}$~= 0.9 (a) and 2.36 TeV (b) are fitted with
two analytic functions. The dashed curves represent the fit
results of the three-component function [Eq. (3)], while the
dotted curves represent the fit results of the $q$-dual function
[Eq. (5)].} \label{fig6}
\end{figure*}

The fit results using the three-component function [Eq. (3), the
dashed curves] and the $q$-dual function [Eq. (5), the dotted
curves] for the CMS data and MC models for the $p_{\rm T}$ spectra
of charged hadrons in the region of $|\eta|<2.4$ in $pp$
collisions at $\sqrt{s}$~= 0.9 (a) and 2.36 TeV (b) are shown in
Figure \ref{fig6}. The results are re-scaled to better visualize
the fit results. The parameters extracted using the two analytic
functions are shown in Tables 1 and 2 at $\sqrt{s}$~= 0.9 TeV and
Tables 3 and 4 at $\sqrt{s}$~= 2.36 TeV as the last row in each case
of data and models.  From Figure 6 one can see that the curves
(functions) fit the data and models predictions well.

From Tables 1 and 3 one can see that the values of $T_B$ of Pythia
model are close to experimental data while QGSJET has higher and
HIJING has lower values than the data. For the values of $N_0$,
all models have lower values than the data at $\sqrt{s}$~= 0.9 TeV
with Pythia and HIJING closer values than QGSJET, while at
$\sqrt{s}$~= 2.36 TeV both Pythia and HIJING have much closer values
to the data than QGSJET. 

From Tables 2 and 4 one can see that all models have higher values
of $T_q$ than the data in case of $\sqrt{s}$~= 2.36 TeV with HIJING
having comparable value. The value of $q$ in data at $\sqrt{s}$~= 0.9
TeV is $1.113\pm0.002$ while at $\sqrt{s}$~= 2.36 is $1.141\pm0.002$.
At $\sqrt{s}$~= 0.9 TeV, the values of Pythia is closer to the data
while HIJING and QGSJET have higher value of $q$. The kinetic
freeze-out volume $V$ in case of data at $\sqrt{s}$~= 0.9 TeV is
$(1.162\pm0.056)\times10^{4}$ while at $\sqrt{s}$~= 2.36 TeV it is
equal to $(9.995\pm0.229)\times10^{4}$. Agian Pythia has closer
value at $\sqrt{s}$~= 0.9 TeV then the HIJING and QGSJET which have
higher and lower values respectively while in case of $\sqrt{s}$~=
2.36 TeV, all models have lower values than the data.

\begin{figure*}[!ht]
\centering
\includegraphics[width=0.95\textwidth]{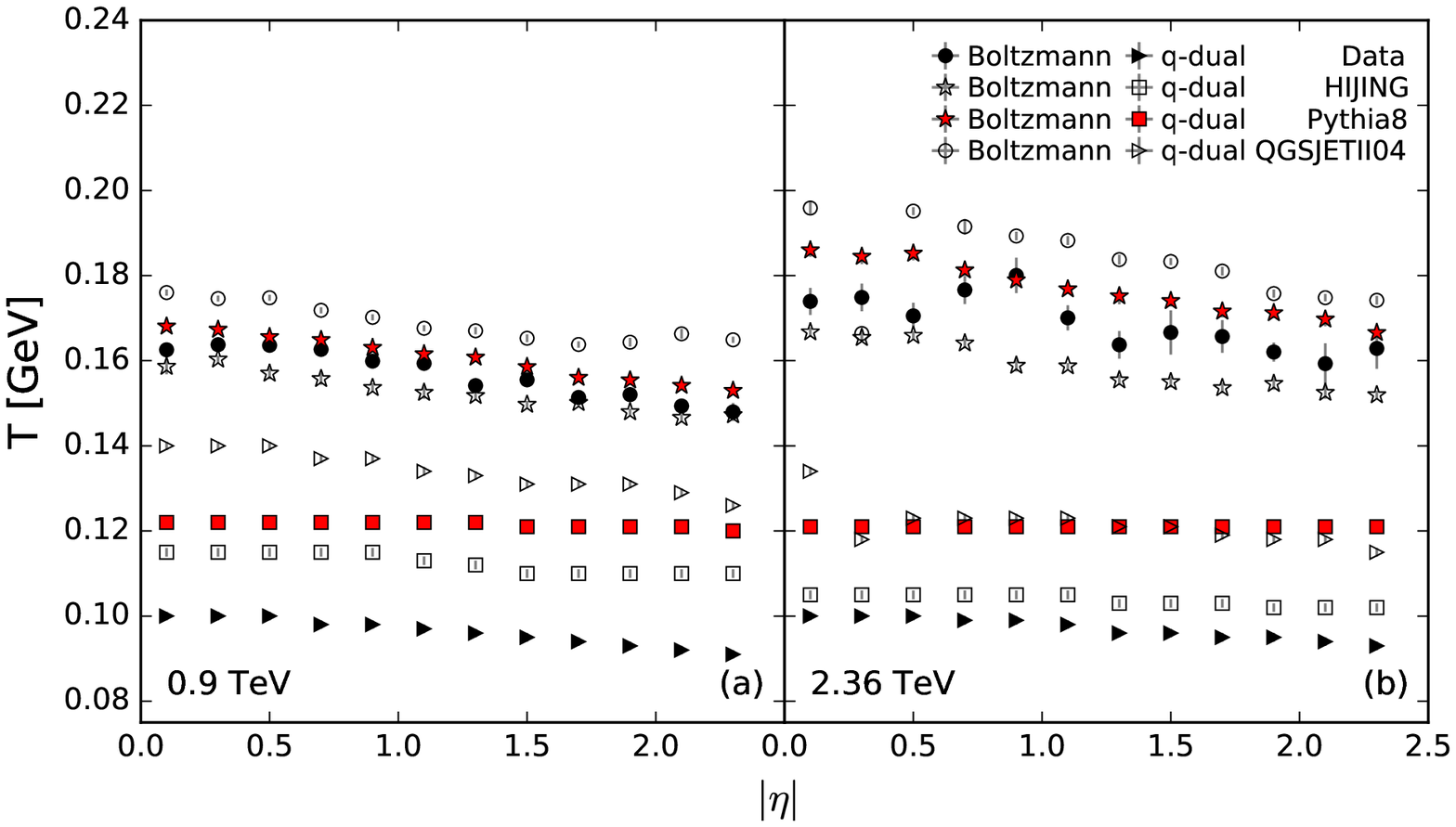}
\caption{Correlations of the effective temperature $T$ ($T_B$ or
$T_q$) with $|\eta|$ of charged hadrons extracted from the three-component function [Eq. (3), the Boltzmann distribution related] and the $q$-dual function [Eq. (5)] for the data as well
as the MC model predictions. In most cases, the error bars for MC models are smaller than symbol size.} \label{fig7}
\end{figure*}

To see the variation of the effective temperature $T$ ($T_B$ and $T_q$), Figure~\ref{fig7} shows the values of $T$ extracted from the three-component function [Eq. (3)] and the $q$-dual function
[Eq. (5)] as a function of $|\eta|$ at $\sqrt{s}$ = 0.9 (a) and 2.36 TeV (b). It is clear from the graphs that the values of $T_B$ is larger compared to $T_q$ as the latter is affected by $q$. In fact, the effect of a larger $T_q$ can be covered by a larger $q$. That is, if $q>1$, one can obtain a smaller $T_q$; while if $q\rightarrow1$, one can obtain a larger $T_q$.

From 0.9 to 2.36 TeV, $T_B$ increases. Meanwhile, $T_B$ decreases
with increasing $|\eta|$. It is also clear from the graphs that
the values of $T_B$ for Pythia are closer to that for the experimental data while QGSJET has higher and HIJING has a lower value. From 0.9 to 2.36 TeV, $T_q$ increases slightly for the data
and is almost invariant for the MC models in general. Meanwhile,
$T_q$ is a very slowly varying function of $|\eta|$ but generally
has a decreasing trend with increasing $|\eta|$. The value of
$T_q$ extracted from the HIJING prediction is closer to that from
the experimental data, though the HIJING overestimates it. The other
two predictions overestimate $T_q$ to a higher degree.

%Figure \ref{fig8} shows the pseudorapidity ($\eta$) distribution of charged particles in $pp$ collisions at $\sqrts$~= 0.9 TeV (left) and $\sqrts$~= 2.36 TeV (right) measured by CMS experiment \cite{cms} in comparison with the models predicton including HIJING, Pythia and QGSJET at both center of mass energies. The Pythia model reproduced the experimental data very well and over the entire $\eta$ region. The other two models underpredict the $\eta$ distribution with HIJING having better comparison than the QGSJET. Furthermore, with an increase in the center of mass energy, the HIJING prediction got better while the QGSJET got worst. The lower panel of the graph shows the ratio of the Monte Carlo prediction to the data.

%\begin{figure*}[!ht]
%\centering
%\includegraphics[width=0.49\textwidth]{14a.pdf}
%\includegraphics[width=0.49\textwidth]{14b.pdf}
%\caption{The experimental measurements as well as models predictions of the transverse momentum spectra shown in one wider $\eta$ bins (0 $<$ $\eta$ $<$ 2.4) of unidentified charge hadrons in $pp$ collision at \sqrts~=0.9 TeV (left) and \sqrts~=2.36 TeV (right).}
%\label{fig8}
%\end{figure*}

%\begin{sidewaystable}
\begin{table*}[!htbp]
{\small Table 3. Values of free parameter $T_B$, normalization
constant $N_0$, $\chi^2$ and ndof extracted from the $p_{\rm T}$
spectra of charged hadrons produced in $pp$ collisions at
$\sqrt{s}$~=2.36 TeV using the three-component function.

\begin{center}
\begin{tabular} {cccccccccccc}\\ \hline\hline
Case & $\eta$ & $T_B$ (GeV) & $N_0$ & $\chi^2$/ndof \\
\hline
$         $  & $0.1         $  & $0.174\pm0.003$ & $(9.132\pm0.139)\times10^{-1}$  & $2/8$\\
$         $  & $0.3         $  & $0.175\pm0.003$ & $(9.210\pm0.119)\times10^{-1}$  & $1/8$\\
$         $  & $0.5         $  & $0.171\pm0.003$ & $(9.076\pm0.159)\times10^{-1}$  & $2/8$\\
$         $  & $0.7         $  & $0.177\pm0.003$ & $(9.496\pm0.139)\times10^{-1}$  & $2/8$\\
$         $  & $0.9         $  & $0.180\pm0.004$ & $(9.657\pm0.013)\times10^{-1}$  & $1/8$\\
Data         & $1.1         $  & $0.170\pm0.003$ & $(9.253\pm0.119)\times10^{-1}$  & $1/8$\\
$         $  & $1.3         $  & $0.164\pm0.003$ & $(9.456\pm0.119)\times10^{-1}$  & $2/8$\\
$         $  & $1.5         $  & $0.167\pm0.005$ & $(9.554\pm0.299)\times10^{-1}$  & $5/8$\\
$         $  & $1.7         $  & $0.166\pm0.004$ & $(9.788\pm0.159)\times10^{-1}$  & $2/8$\\
$         $  & $1.9         $  & $0.162\pm0.002$ & $(9.716\pm0.119)\times10^{-1}$  & $1/8$\\
$         $  & $2.1         $  & $0.159\pm0.005$ & $(9.810\pm0.319)\times10^{-1}$  & $4/8$\\
$         $  & $2.3         $  & $0.163\pm0.005$ & $(9.543\pm0.199)\times10^{-1}$  & $5/8$\\
$         $  & $|\eta| <2.4 $  & $0.167\pm0.002$ & $(2.267\pm0.045)\times10^{1}$   & $12/18$\\
\hline
$         $  & $0.1         $  & $0.167\pm0.010$ & $(8.311\pm0.059)\times10^{-1}$  & $1/8$\\
$         $  & $0.3         $  & $0.165\pm0.010$ & $(8.392\pm0.049)\times10^{-1}$  & $1/8$\\
$         $  & $0.5         $  & $0.166\pm0.010$ & $(8.468\pm0.049)\times10^{-1}$  & $1/8$\\
$         $  & $0.7         $  & $0.164\pm0.012$ & $(8.629\pm0.051)\times10^{-1}$  & $1/8$\\
$         $  & $0.9         $  & $0.159\pm0.014$ & $(8.715\pm0.059)\times10^{-1}$  & $1/8$\\
HIJING       & $1.1         $  & $0.159\pm0.007$ & $(8.813\pm0.019)\times10^{-1}$  & $1/8$\\
$         $  & $1.3         $  & $0.155\pm0.012$ & $(8.955\pm0.049)\times10^{-1}$  & $1/8$\\
$         $  & $1.5         $  & $0.155\pm0.013$ & $(9.015\pm0.065)\times10^{-1}$  & $1/8$\\
$         $  & $1.7         $  & $0.154\pm0.013$ & $(9.075\pm0.039)\times10^{-1}$  & $2/8$\\
$         $  & $1.9         $  & $0.155\pm0.014$ & $(9.092\pm0.049)\times10^{-1}$  & $2/8$\\
$         $  & $2.1         $  & $0.153\pm0.015$ & $(9.056\pm0.027)\times10^{-1}$  & $2/8$\\
$         $  & $2.3         $  & $0.152\pm0.010$ & $(9.015\pm0.053)\times10^{-1}$  & $3/8$\\
$         $  & $|\eta| <2.4 $  & $0.157\pm0.010$ & $(2.110\pm0.063)\times10^{1}$   & $27/24$\\
\hline
$         $  & $0.1         $  & $0.186\pm0.015$ & $(8.763\pm0.079)\times10^{-1}$  & $1/8$\\
$         $  & $0.3         $  & $0.184\pm0.018$ & $(8.864\pm0.079)\times10^{-1}$  & $2/8$\\
$         $  & $0.5         $  & $0.185\pm0.016$ & $(8.921\pm0.029)\times10^{-1}$  & $1/8$\\
$         $  & $0.7         $  & $0.181\pm0.012$ & $(9.045\pm0.053)\times10^{-1}$  & $1/8$\\
$         $  & $0.9         $  & $0.179\pm0.017$ & $(9.166\pm0.079)\times10^{-1}$  & $2/8$\\
Pythia8      & $1.1         $  & $0.177\pm0.011$ & $(9.288\pm0.043)\times10^{-1}$  & $1/8$\\
$         $  & $1.3         $  & $0.175\pm0.019$ & $(9.407\pm0.047)\times10^{-1}$  & $2/8$\\
$         $  & $1.5         $  & $0.174\pm0.014$ & $(9.467\pm0.067)\times10^{-1}$  & $3/8$\\
$         $  & $1.7         $  & $0.172\pm0.014$ & $(9.487\pm0.039)\times10^{-1}$  & $2/8$\\
$         $  & $1.9         $  & $0.171\pm0.012$ & $(9.527\pm0.075)\times10^{-1}$  & $2/8$\\
$         $  & $2.1         $  & $0.170\pm0.016$ & $(9.409\pm0.059)\times10^{-1}$  & $2/8$\\
$         $  & $2.3         $  & $0.167\pm0.016$ & $(9.332\pm0.046)\times10^{-1}$  & $3/8$\\
$         $  & $|\eta| <2.4 $  & $0.171\pm0.023$ & $(2.216\pm0.054)\times10^{1}$   & $9/18$\\
\hline
$           $  & $0.1         $  & $0.196\pm0.001$ & $(6.863\pm0.019)\times10^{-1}$  & $0.5/8$\\
$           $  & $0.3         $  & $0.166\pm0.001$ & $(8.390\pm0.047)\times10^{-1}$  & $1/8$\\
$           $  & $0.5         $  & $0.195\pm0.001$ & $(7.001\pm0.023)\times10^{-1}$  & $0.2/8$\\
$           $  & $0.7         $  & $0.191\pm0.002$ & $(7.084\pm0.015)\times10^{-1}$  & $0.4/8$\\
$           $  & $0.9         $  & $0.189\pm0.001$ & $(7.184\pm0.011)\times10^{-1}$  & $0.4/8$\\
QGSJETII-04    & $1.1         $  & $0.188\pm0.001$ & $(7.302\pm0.039)\times10^{-1}$  & $1/8$\\
$           $  & $1.3         $  & $0.184\pm0.001$ & $(7.347\pm0.015)\times10^{-1}$  & $1/8$\\
$           $  & $1.5         $  & $0.183\pm0.001$ & $(7.423\pm0.010)\times10^{-1}$  & $2/8$\\
$           $  & $1.7         $  & $0.181\pm0.001$ & $(7.425\pm0.025)\times10^{-1}$  & $2/8$\\
$           $  & $1.9         $  & $0.176\pm0.001$ & $(7.432\pm0.023)\times10^{-1}$  & $2/8$\\
$           $  & $2.1         $  & $0.175\pm0.001$ & $(7.392\pm0.031)\times10^{-1}$  & $2/8$\\
$           $  & $2.3         $  & $0.174\pm0.001$ & $(7.350\pm0.045)\times10^{-1}$  & $2/8$\\
$           $  & $|\eta| <2.4 $  & $0.183\pm0.002$ & $(1.741\pm0.006)\times10^{1}$   & $5/18$\\
\hline

\end{tabular}%
\end{center}}
\end{table*}
%\end{sidewaystable}

%\begin{sidewaystable}
\begin{table*}[!htbp]
{\small Table 4. Values of free parameters $T_q$ and $q$, kinetic
freeze-out volume $V$, $\chi^2$ and ndof extracted from the
$p_{\rm T}$ spectra of charged hadrons produced in $pp$ collisions
at $\sqrt{s}$~=2.36 TeV using the $q$-dual function.

\begin{center}
\begin{tabular} {cccccccccccc}\\ \hline\hline
Case & $\eta$ & $T_q$ (GeV) & $q$  & $V$ (fm$^3$) & $\chi^2$/ndof \\
\hline
$         $  & $0.1         $  & $0.100\pm0.001$ & $1.145\pm0.001$  &$(5.397\pm0.072)\times10^{4}$  & $2/11$\\
$         $  & $0.3         $  & $0.100\pm0.001$ & $1.148\pm0.001$  &$(5.315\pm0.098)\times10^{4}$  & $2/11$\\
$         $  & $0.5         $  & $0.100\pm0.001$ & $1.147\pm0.002$  &$(5.287\pm0.143)\times10^{4}$  & $3/11$\\
$         $  & $0.7         $  & $0.099\pm0.001$ & $1.147\pm0.002$  &$(5.605\pm0.091)\times10^{4}$  & $2/11$\\
$         $  & $0.9         $  & $0.099\pm0.001$ & $1.151\pm0.003$  &$(5.568\pm0.145)\times10^{4}$  & $4/11$\\
Data         & $1.1         $  & $0.098\pm0.001$ & $1.149\pm0.002$  &$(5.660\pm0.121)\times10^{4}$  & $3/11$\\
$         $  & $1.3         $  & $0.096\pm0.001$ & $1.148\pm0.002$  &$(6.236\pm0.135)\times10^{4}$  & $3/11$\\
$         $  & $1.5         $  & $0.096\pm0.001$ & $1.148\pm0.003$  &$(6.264\pm0.203)\times10^{4}$  & $6/11$\\
$         $  & $1.7         $  & $0.095\pm0.001$ & $1.148\pm0.002$  &$(6.471\pm0.146)\times10^{4}$  & $4/11$\\
$         $  & $1.9         $  & $0.095\pm0.001$ & $1.148\pm0.002$  &$(6.562\pm0.137)\times10^{4}$  & $3/11$\\
$         $  & $2.1         $  & $0.094\pm0.001$ & $1.139\pm0.003$  &$(7.110\pm0.210)\times10^{4}$  & $4/11$\\
$         $  & $2.3         $  & $0.093\pm0.001$ & $1.144\pm0.003$  &$(6.990\pm0.219)\times10^{4}$  & $7/11$\\
$         $  & $|\eta| <2.4 $  & $0.115\pm0.001$ & $1.141\pm0.002$  &$(9.995\pm0.229)\times10^{4}$  & $26/21$\\
\hline
$         $  & $0.1         $  & $0.105\pm0.001$ & $1.118\pm0.002$  &$(4.725\pm0.132)\times10^{4}$  & $22/11$\\
$         $  & $0.3         $  & $0.105\pm0.001$ & $1.118\pm0.002$  &$(4.767\pm0.145)\times10^{4}$  & $20/11$\\
$         $  & $0.5         $  & $0.105\pm0.001$ & $1.116\pm0.001$  &$(4.854\pm0.145)\times10^{4}$  & $20/11$\\
$         $  & $0.7         $  & $0.105\pm0.001$ & $1.114\pm0.001$  &$(4.981\pm0.161)\times10^{4}$  & $21/11$\\
$         $  & $0.9         $  & $0.105\pm0.001$ & $1.112\pm0.001$  &$(5.095\pm0.128)\times10^{4}$  & $15/11$\\
HIJING       & $1.1         $  & $0.105\pm0.001$ & $1.112\pm0.002$  &$(5.112\pm0.131)\times10^{4}$  & $17/11$\\
$         $  & $1.3         $  & $0.103\pm0.001$ & $1.113\pm0.002$  &$(5.595\pm0.128)\times10^{4}$  & $14/11$\\
$         $  & $1.5         $  & $0.103\pm0.001$ & $1.112\pm0.002$  &$(5.559\pm0.096)\times10^{4}$  & $13/11$\\
$         $  & $1.7         $  & $0.103\pm0.001$ & $1.112\pm0.001$  &$(5.608\pm0.094)\times10^{4}$  & $14/11$\\
$         $  & $1.9         $  & $0.102\pm0.001$ & $1.112\pm0.002$  &$(5.801\pm0.096)\times10^{4}$  & $15/11$\\
$         $  & $2.1         $  & $0.102\pm0.001$ & $1.112\pm0.002$  &$(5.814\pm0.100)\times10^{4}$  & $14/11$\\
$         $  & $2.3         $  & $0.102\pm0.001$ & $1.112\pm0.002$  &$(5.814\pm0.103)\times10^{4}$  & $14/11$\\
$         $  & $|\eta| <2.4 $  & $0.120\pm0.001$ & $1.127\pm0.002$  &$(1.724\pm0.057)\times10^{4}$  & $108/27$\\
\hline
$         $  & $0.1         $  & $0.121\pm0.001$ & $1.111\pm0.002$  &$(3.226\pm0.132)\times10^{4}$  & $23/11$\\
$         $  & $0.3         $  & $0.121\pm0.001$ & $1.110\pm0.002$  &$(3.275\pm0.129)\times10^{4}$  & $23/11$\\
$         $  & $0.5         $  & $0.121\pm0.001$ & $1.109\pm0.003$  &$(3.313\pm0.125)\times10^{4}$  & $24/11$\\
$         $  & $0.7         $  & $0.121\pm0.001$ & $1.107\pm0.002$  &$(3.352\pm0.104)\times10^{4}$  & $20/11$\\
$         $  & $0.9         $  & $0.121\pm0.001$ & $1.104\pm0.002$  &$(3.439\pm0.112)\times10^{4}$  & $25/11$\\
Pythia8      & $1.1         $  & $0.121\pm0.001$ & $1.104\pm0.002$  &$(3.488\pm0.101)\times10^{4}$  & $20/11$\\
$         $  & $1.3         $  & $0.121\pm0.001$ & $1.101\pm0.002$  &$(3.579\pm0.112)\times10^{4}$  & $22/11$\\
$         $  & $1.5         $  & $0.121\pm0.001$ & $1.100\pm0.002$  &$(3.611\pm0.122)\times10^{4}$  & $21/11$\\
$         $  & $1.7         $  & $0.121\pm0.001$ & $1.098\pm0.002$  &$(3.528\pm0.102)\times10^{4}$  & $30/11$\\
$         $  & $1.9         $  & $0.121\pm0.001$ & $1.096\pm0.003$  &$(3.662\pm0.112)\times10^{4}$  & $36/11$\\
$         $  & $2.1         $  & $0.121\pm0.001$ & $1.096\pm0.003$  &$(3.631\pm0.111)\times10^{4}$  & $29/11$\\
$         $  & $2.3         $  & $0.121\pm0.001$ & $1.094\pm0.003$  &$(3.623\pm0.115)\times10^{4}$  & $29/11$\\
$         $  & $|\eta| <2.4 $  & $0.132\pm0.001$ & $1.121\pm0.001$  &$(1.410\pm0.050)\times10^{4}$  & $40/21$\\
\hline
$           $  & $0.1         $  & $0.134\pm0.001$ & $1.102\pm0.002$  &$(1.891\pm0.064)\times10^{4}$  & $21/11$\\
$           $  & $0.3         $  & $0.118\pm0.001$ & $1.098\pm0.002$  &$(3.473\pm0.122)\times10^{4}$  & $38/11$\\
$           $  & $0.5         $  & $0.123\pm0.001$ & $1.117\pm0.003$  &$(2.379\pm0.033)\times10^{4}$  & $17/11$\\
$           $  & $0.7         $  & $0.123\pm0.001$ & $1.117\pm0.002$  &$(2.399\pm0.031)\times10^{4}$  & $13/11$\\
$           $  & $0.9         $  & $0.123\pm0.001$ & $1.115\pm0.001$  &$(2.450\pm0.045)\times10^{4}$  & $10/11$\\
QGSJETII-04    & $1.1         $  & $0.123\pm0.001$ & $1.114\pm0.001$  &$(2.489\pm0.046)\times10^{4}$  & $10/11$\\
$           $  & $1.3         $  & $0.121\pm0.001$ & $1.114\pm0.001$  &$(2.663\pm0.034)\times10^{4}$  & $8/11$\\
$           $  & $1.5         $  & $0.121\pm0.001$ & $1.113\pm0.002$  &$(2.682\pm0.059)\times10^{4}$  & $10/11$\\
$           $  & $1.7         $  & $0.119\pm0.001$ & $1.113\pm0.002$  &$(2.849\pm0.049)\times10^{4}$  & $9/11$\\
$           $  & $1.9         $  & $0.118\pm0.001$ & $1.113\pm0.002$  &$(2.922\pm0.051)\times10^{4}$  & $8/11$\\
$           $  & $2.1         $  & $0.118\pm0.001$ & $1.112\pm0.001$  &$(2.909\pm0.050)\times10^{4}$  & $9/11$\\
$           $  & $2.3         $  & $0.115\pm0.001$ & $1.113\pm0.002$  &$(3.160\pm0.059)\times10^{4}$  & $9/11$\\
$           $  & $|\eta| <2.4 $  & $0.125\pm0.001$ & $1.141\pm0.001$  &$(1.246\pm0.045)\times10^{4}$  & $78/21$\\
\hline
\end{tabular}%
\end{center}}
\end{table*}
\section{Conclusion}\label{sec4}

We have presented simulation results of the transverse momentum
{\ppt} spectra of charged hadrons in small bins of pseudorapidity
$|\eta|$ with a gap of 0.2 units from 0 to 2.4 at {\ppt} from 0.1
to 2 GeV/$c$ and a single bin of $|\eta|<2.4$ for {\ppt} from 0.1
to 4 GeV/$c$ in $pp$ collisions at $\sqrt{s}$ = 0.9 and 2.36 TeV. The
predictions of different Monte Carlo event generators including
HIJING, Pythia, and QGSJETII are used in comparison with the
experimental data measured by the CMS Collaboration at the same
energies. The spectra measured by the CMS Collaboration were
normalized to all NSD events. The same initial conditions are used in the case of event generators as the data. The Pythia model reproduced
the {\ppt} distributions at different $|\eta|$ bins in most of the
{\ppt} range. Furthermore, the model depicts better results in
case of the $|\eta|<2.4$ than HIJING and QGSJETII as well. The
later two models reproduce the {\ppt} distribution in a limited
range of the {\ppt} only. Overall, the HIJING prediction is found
to be better than the QGSJETII whose prediction got better with
increase in the energy from 0.9 to 2.36 TeV.

Furthermore, to analyse the {\ppt} spectra of charged hadrons
measured by the CMS experiment and generated from the three MC
models in $pp$ collisions at high energy, we used a
three-component function structured from the Boltzmann
distribution and the $q$-dual function from the $q$-dual
statistics to extract parameter values relevant for the study of
bulk properties of nuclear matter. The effective temperatures
extracted from the experimental data based on the two functions
are found to be decreasing with $|\eta|$ but slightly increasing
with $\sqrt{s}$. The normalization constant $N_0$ and the kinetic
freeze-out volume $V$ increase with $|\eta|$ and $\sqrt{s}$. We have
also applied the two functions over the MC model predictions and
found that the extracted parameters follow the same trend as
explained above for the data in most cases. Due to the influence
of $q$, $T_q$ shows almost energy independent for the cases of MC
models. The values of the parameters extracted from the HIJING and
Pythia models are closer to the experimental data than the
QGSJETII model supporting that the HIJING and Pythia models
predict the data better as compared to the QGSJETII model.
\\
\\
\\
{\bf Acknowledgments}

The work of Shanxi Group was supported by the National Natural
Science Foundation of China under Grant Nos. 12147215, 12047571,
11575103, and 11947418, the Scientific and Technological
Innovation Programs of Higher Education Institutions in Shanxi
(STIP) under Grant No. 201802017, the Shanxi Provincial Natural
Science Foundation under Grant No. 201901D111043, and the Fund for
Shanxi ``1331 Project" Key Subjects Construction.
\\

\bibliographystyle{unsrt}
\bibliography{bibliography.bib}

\end{multicols}
\end{document}